\documentclass{aa}  
\usepackage{graphicx}
\usepackage{txfonts}
\usepackage{natbib}
\begin{document}
\title{New triple systems in the RasTyc sample of \\
stellar X-ray sources
\thanks{Based on observations collected at the \textsl{Observatoire de Haute Provence} (France) and 
	the \textsl{M. G. Fracastoro} station (Serra La Nave, Mt. Etna, 1750\,m a.s.l.) of the Catania 
	Astrophysical Observatory (Italy)}
\fnmsep\thanks{Tables 2 to 4 are available in electronic form at the CDS via anonymous FTP to 
		   {\tt cdsarc.u-strasbg.fr (130.79.128.5)} or via 
		   {\tt http://cdsweb.u-strasbg.fr/cgi-bin/qcat?J/A+A/} as well as via 
		   {\tt http://www.edpsciences.org/}}
\fnmsep\thanks{Figures 1\,--\,4 and 8 are only available in electronic form via 
		    {\tt http://www.edpsciences.org/}}
	}

\author{A. Klutsch\inst{1} \and A. Frasca\inst{2} \and P. Guillout\inst{1} \and 
	R. Freire Ferrero\inst{1} \and E. Marilli\inst{2} \and 
	G. Mignemi\inst{1}\fnmsep\inst{3} \and K. Biazzo\inst{2}\fnmsep\inst{4}}

\offprints{A. Klutsch \\
\email{klutsch@astro.u-strasbg.fr}\\}

\institute{Observatoire Astronomique, Universit\'e de Strasbourg \& CNRS (UMR 7550), 11 rue de l'Universit\'e, 
	   67000 Strasbourg, France
     \and
	   INAF - Osservatorio Astrofisico di Catania, via S. Sofia, 78, 95123 Catania, Italy
     \and
	   Dipartimento di Fisica e Astronomia, Universit\`a di Catania, via S. Sofia, 78, 95123 Catania, Italy
     \and
	   ESO - European Southern Observatory, Karl-Schwarzschild-Str. 3, 85748 Garching bei M\"unchen, Germany
	   }

\date{Received 6 May 2008 / Accepted 22 July 2008}
 
\abstract 
{ During the study of a large set of late-type stellar X-ray sources, we discovered a large fraction 
of multiple systems.} 
{ In this paper we investigate the orbital elements and kinematic properties of three new 
spectroscopic triple systems as well as spectral types and astrophysical parameters ($T_{\rm eff}$, $\log g$, 
$v\sin i$, $\log N$(Li)) of their components.}
{ We conducted follow-up optical observations, both photometric and spectroscopic at high 
resolution, of these systems. We used a synthetic approach 
and the cross-correlation method to derive most of the stellar parameters.}
{ We estimated reliable radial velocities and deduced the orbital elements  
of the inner binaries. The comparison of the observed spectra with synthetic composite ones, 
obtained as the weighted sum of three spectra of non-active reference stars, allowed us to determine 
the stellar parameters for each component of these systems. We found all are only composed of  
main sequence stars.}
{ These three systems are certainly stable hierarchical triples composed of short-period inner binaries 
plus a tertiary component in a long-period orbit. From their kinematics and/or Lithium content, these 
systems result to be fairly young.}
\titlerunning{New triple systems in the RasTyc sample of stellar X-ray sources}
\authorrunning{Klutsch et al.}
\maketitle

\section{Introduction}
\label{sec:Intro}

Binary and multiple stars are very important astrophysical laboratories. In particular, spectro-photometric 
and spectro-astrometric binaries offer the unique opportunity to determine, with a high level of accuracy, 
the basic stellar parameters (mass, radius, and effective temperature) to study stellar structure and 
evolution. However, the formation and evolution of binary stars are still debated subjects
\citep[e.g.,][]{ZM2001}. Especially, a still unsolved problem is the formation of close binaries
with main sequence components separated by few solar radii that in the proto-stellar phase should 
have been in contact.

In the last years, to answer many of these open questions, relevant observational and theoretical efforts 
are being done to improve continuously the statistics of binary systems with different periods, mass ratios, 
etc. \citep[e.g.,][]{Tok06}. 

\begin{table*}
\caption[RasTyc sources]{Main data of the three {\it RasTyc} sources from the literature.}
\begin{center}
\begin{tabular}{clccccrrcc}
\hline
\hline
\noalign{\medskip}
\textsl{RasTyc} Name & Name & $\alpha$ (2000) & $\delta$ (2000) & V$_T^{\,\,\rm a}$  & $\pi^{\,\,\rm b}$ & $\mu_{\alpha}^{\,\,\rm a}$~~~ & $\mu_{\delta}^{\,\,\rm a}$~~~ & X-ray source & Counts \\
 & & (h m s) & ($\degr ~\arcmin ~\arcsec$) & (mag) & (mas) & \multicolumn{2}{c}{(mas\,yr$^{-1}$)} & 1RXS & (ct\,s$^{-1}$) \\
\noalign{\medskip}
\hline
\noalign{\medskip}
 RasTyc\,0524+6739 & \object{BD+67\,381}  & 05 24 53.2 & +67 39 39 & 9.065 &  7.8$\pm$9.3	& $-0.5$~~  & 26.7~~    & J052454.0+673939 & 4.21$\times 10^{-1}$ \\
 RasTyc\,1828+3506 & \object{BD+35\,3261} & 18 28 50.3 & +35 06 34 & 9.049 & 12.2$\pm$8.2	& 12.3~~    & $-3.5$~~  & J182849.7+350637 & 6.24$\times 10^{-2}$ \\
 RasTyc\,2034+8253 & \object{BD+82\,622}  & 20 34 27.5 & +82 53 35 & 9.730 &  ---		& 61.5~~    & 35.5~~    & J203426.2+825334 & 3.75$\times 10^{-1}$\\
\hline
\noalign{\medskip}
\end{tabular}
\begin{tabular}{cc}
$^{\rm a}$ $V$ magnitude and proper motions from the TYCHO-2 catalog \citep{Hog00}; & $^{\rm b}$ Parallax from the TYCHO-1 catalog \citep{Hipp} \\
\end{tabular}
\end{center}
\label{Tab:SourcesRasTyc}
\end{table*}

\onlfig{1}{
\begin{figure*}[!th] 
	\centering
	\includegraphics[width=9.0cm]{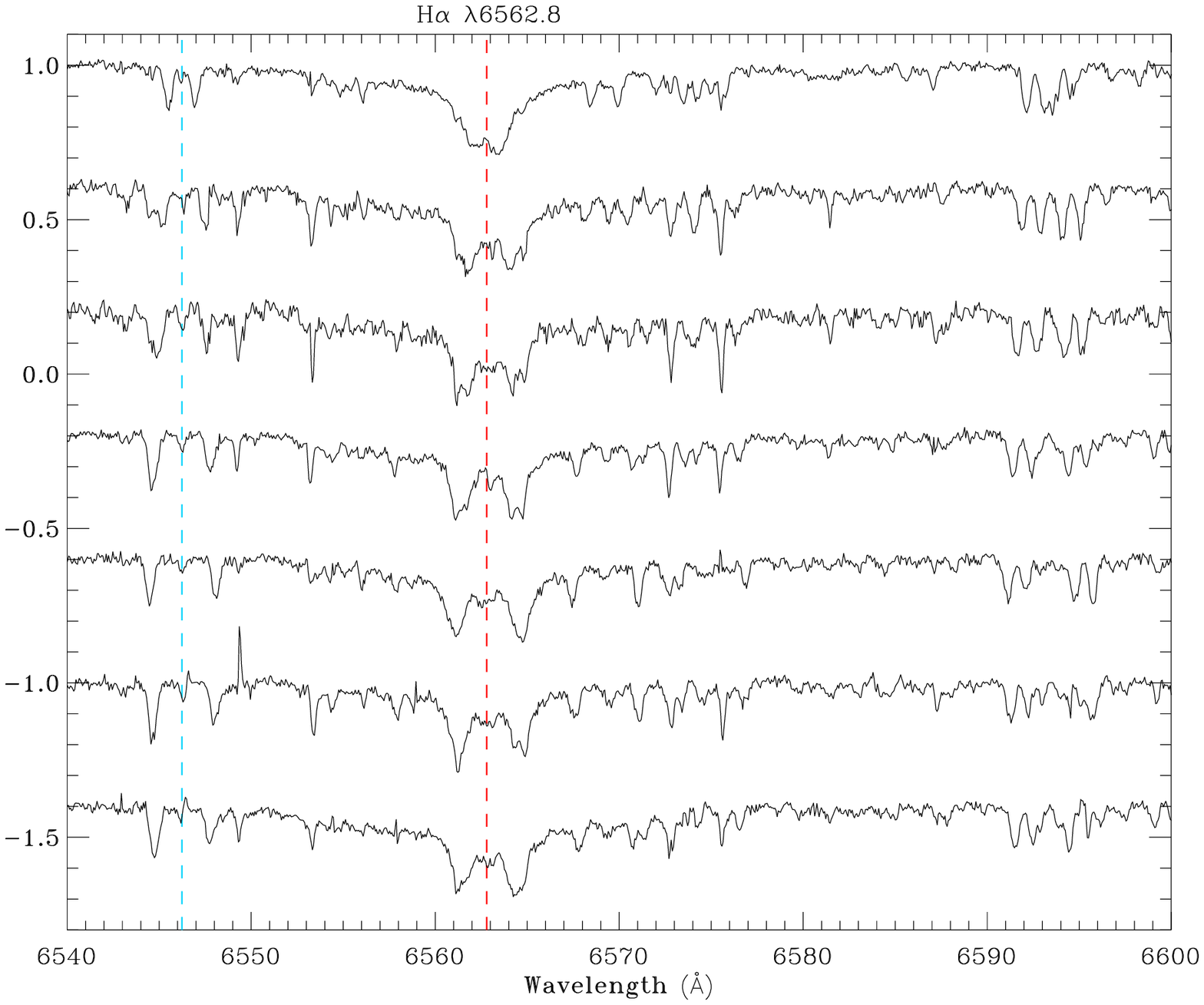}
	\includegraphics[width=9.0cm]{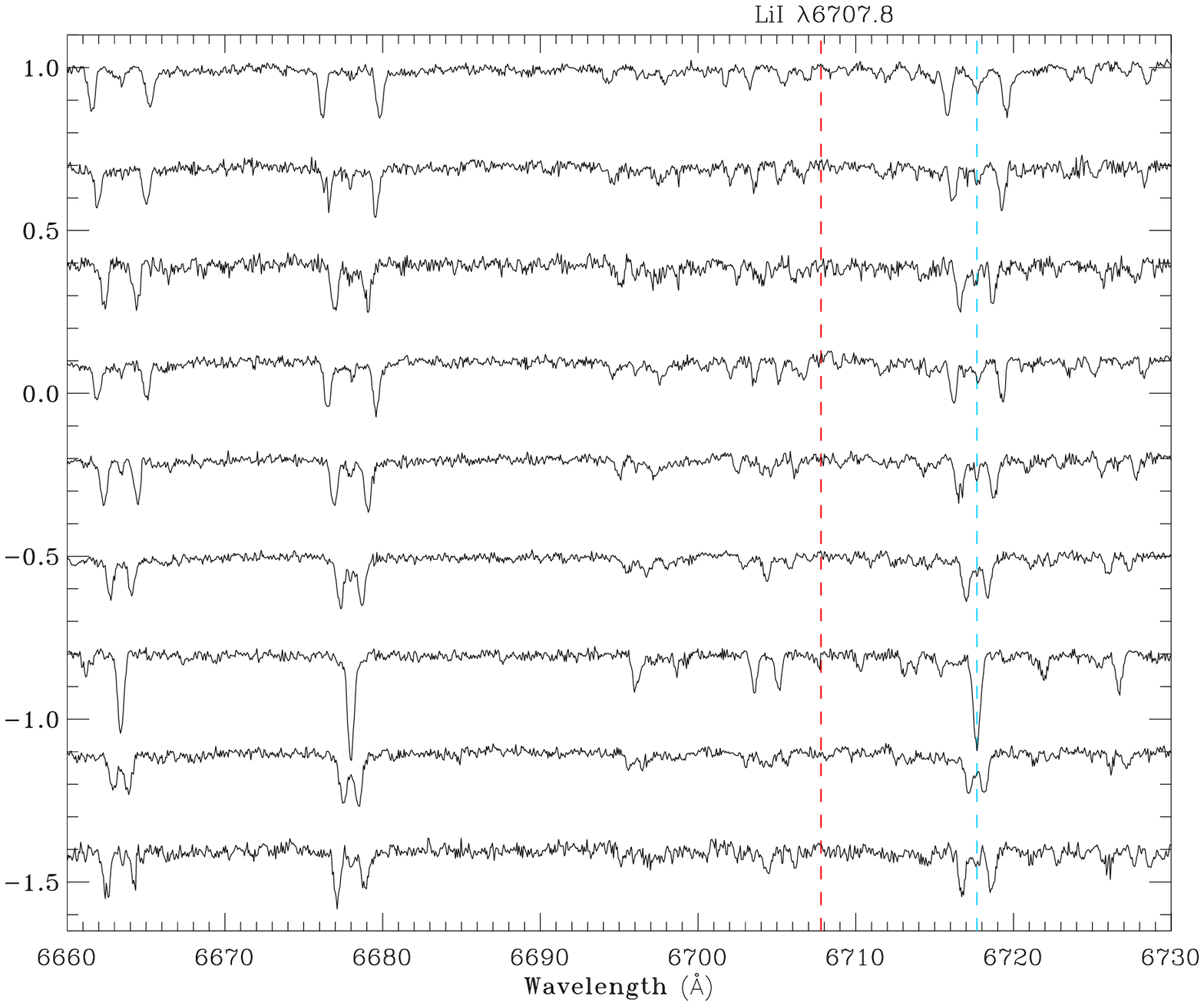}
	\includegraphics[width=9.0cm]{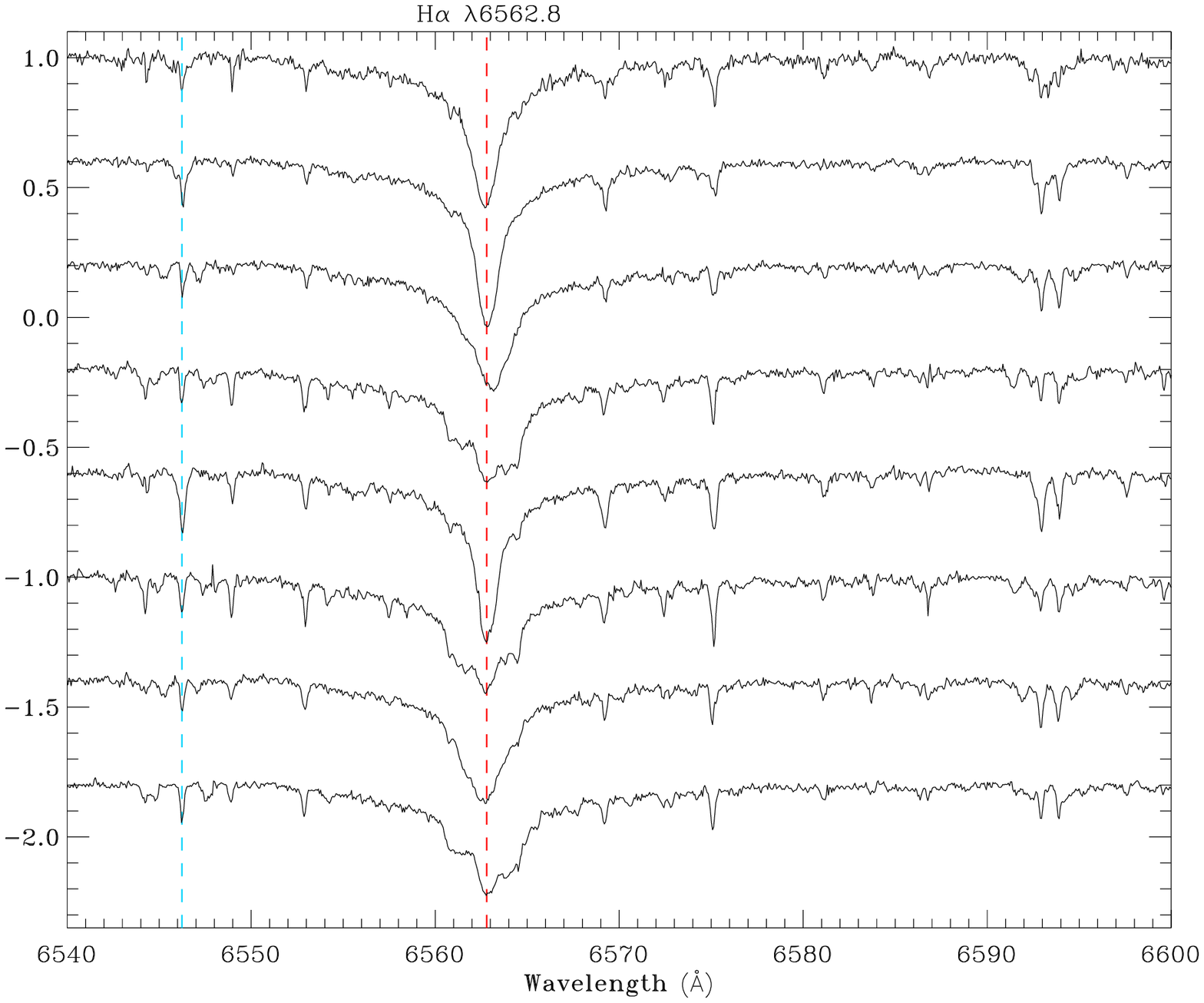}
	\includegraphics[width=9.0cm]{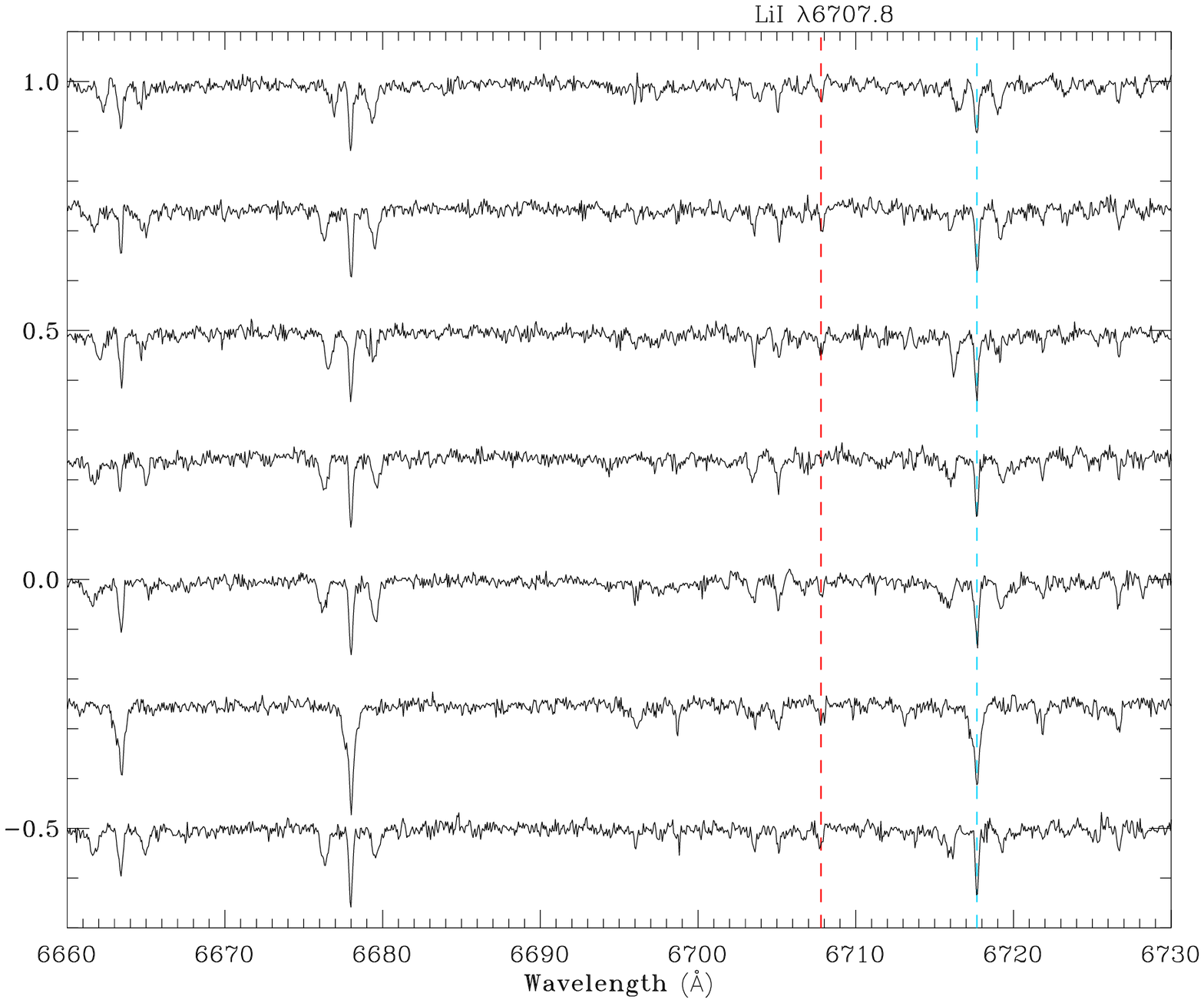}
	\includegraphics[width=9.0cm]{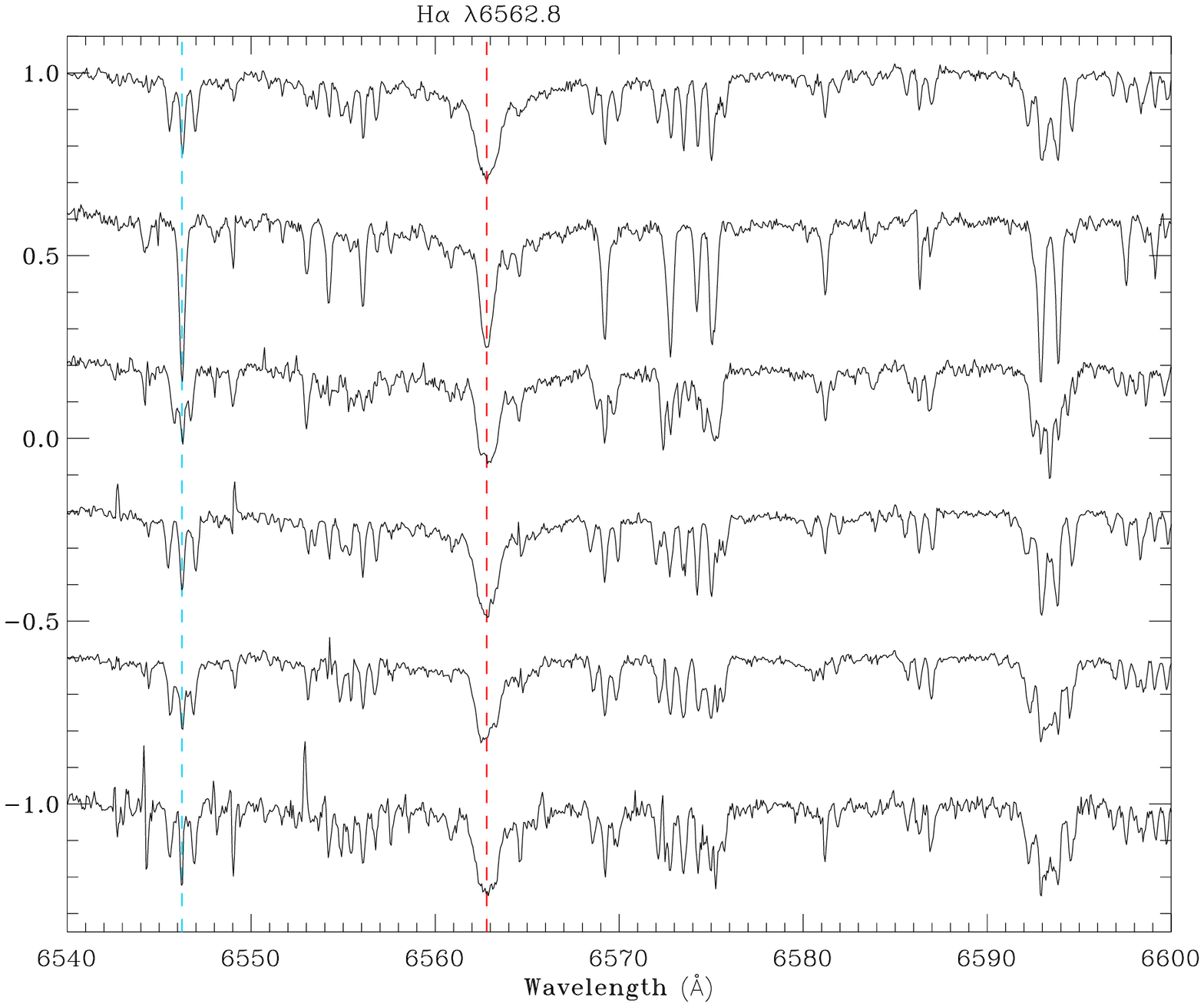}
	\includegraphics[width=9.0cm]{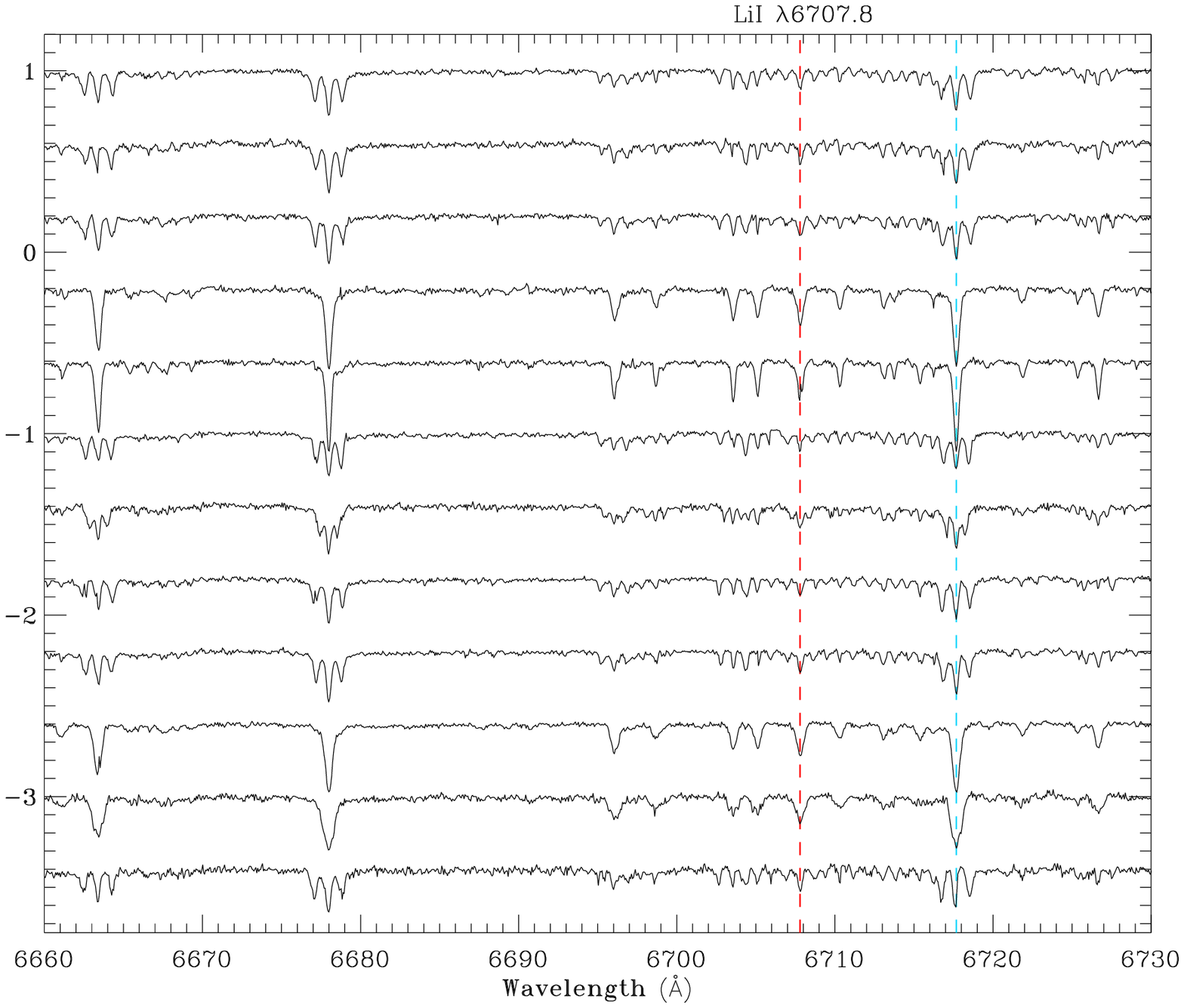}
	\caption{\label{fig:EvolutionSpectraAll} High resolution spectra of RasTyc\,0524+6739 (top panels), 
	RasTyc\,1828+3506 (middle panels) and RasTyc\,2034+8253 (lower panels) acquired with the A{\sc urelie} 
 	spectrograph at the  152-cm telescope of the OHP both in the H$\alpha$ (left panels) and the Lithium spectral 
	regions (right panels). The laboratory wavelengths of the \ion{Fe}{i}\,$\lambda$6546.2 and the H$\alpha$ lines 
	as well as those of the \ion{Li}{i}\,$\lambda$6707.8, and the \ion{Ca}{i}\,$\lambda$6717.7 are marked with 
	vertical dashed lines in the H$\alpha$ and Lithium spectral regions, respectively.}
\end{figure*}
}

Close binaries containing at least one late-type component, such as RS~CVn and BY~Dra 
systems, are objects with the strongest magnetic activity (starspots, plages, flares) induced by a dynamo 
action in the sub-photospheric convection zone. Their strong activity is mainly due to their very fast rotation
(spin-orbit synchronization by tidal forces) and to proximity effects.

X-ray sky surveys performed in recent years have allowed to identify thousands of active late-type stars in 
the field and in open clusters. Follow-up observations of the optical counterparts of X-ray sources 
have led to discover very young stars far from the typical birth sites, i.e. open clusters and stars forming 
regions \citep[e.g.,][]{Wichmann03, Zickgraf05, Torres2006, Guillout2008}, as well as to detect several 
spectroscopic binaries \citep[e.g.,][]{Wichmann03b, Frasca2006}. The knowledge of the incidence of binaries 
and multiple systems in X-ray selected samples of active stars is extremely important to study the 
recent local star formation history.
 
One of the largest ($\sim14\,000$ active stars) and most comprehensive set of stellar X-ray sources in 
the field is the so-called \textsl{RasTyc} sample, which is the result of the cross-correlation of the ROSAT 
All-Sky Survey (RASS) with the TYCHO catalog \citep{Guillout1999}. We began to analyze a representative 
sub-sample of the \textsl{RasTyc} population in the northern hemisphere \citep{Guillout2008} to obtain some 
reliable statistics about the \textsl{RasTyc} stellar characteristics. For this purpose, we led campaigns of 
high-resolution spectroscopic observations, with the E{\sc lodie} \'echelle spectrograph at the 193-cm 
telescope and the A{\sc urelie} spectrograph at the 152-cm telescope of the {\it Observatoire de Haute Provence} (OHP). For all the sources, we performed a detailed analysis of the cross-correlation function 
(CCF) and found that single-lined (SB1), double-lined (SB2), and triple-lined (SB3) spectroscopic 
systems altogether account for more than 35\,\% of the sample. In particular, at least 10 sources are 
clearly identified as triple systems. Our aim is to determine the orbital and physical parameters of these 
systems. For this reason, we have monitored these new multiple systems, both photometrically and 
spectroscopically, with the 91-cm telescope of the {\it Osservatorio Astrofisico di Catania} (OAC).

The majority of the triple systems studied so far are nearby objects and look as visual binaries where 
one componentis a SB2 system. Here we study three of such newly discovered SB3 late-type systems 
for which we obtained enough data. A detailed analysis of the entire sample of stellar X-ray sources and 
of the multiple systems discovered so far is necessary for drawing statistically significant conclusions.
Nevertheless, the properties of these three systems can give us some insights into the typical composition 
of triple systems among stellar X-ray sources. 

The paper is organized as follows. We summarize briefly the observations and their reduction in 
Sect.~\ref{sec:ObsMet}. The determination of radial velocity and physical parameters ($v\sin i$'s, 
Lithium abundances, etc.) are shown in Sect.~\ref{sec:ObsMet} as well. The photometric modulation, 
the spectral composition, and the age estimate are discussed in Sect.~\ref{sec:Discussion}. The 
conclusions are outlined in Sect.~\ref{sec:Conclusions}.

\section{Observations and spectral analysis}
\label{sec:ObsMet}

These three systems belong to the \textsl{RasTyc} sample of stellar X-ray sources \citep{Guillout1999}. 
Their spectra, acquired at different dates, show alternatively a triple-line or a single-line pattern 
(Fig.~\ref{fig:EvolutionSpectraAll})\footnote{Available in electronic form only.}, that is the unambiguous 
signature of SB3 systems. Table \ref{Tab:SourcesRasTyc} lists the most relevant information from the 
literature. In the following, we briefly outline the observations and analysis methods and refer to 
\citet{Guillout2008} for a detailed description.

\subsection{Observations and data reduction}
\label{subsec:Obs}

Spectroscopic observations were first conducted at the OHP between 2001 and 2005 with the A{\sc urelie} 
spectrograph \citep{Gillet1994} within the framework of a key program on the coud\'e 152-cm telescope. 
We used grating \#7, which yields a high spectral resolution, $R=\lambda/\Delta\lambda$, of about $38\,000$ 
in the wavelength range of our observations, i.e. both in the H$\alpha$ (6490-6630\,\AA) and the Lithium 
(6650-6780\,\AA) spectral regions. We obtained 16, 15 and 18 spectra for RasTyc\,0524+6739, 
RasTyc\,1828+3506 and RasTyc\,2034+8253, respectively. The signal-to-noise ratio, S/N, was in the range 
$70$\,--\,$200$. All these spectra were reduced using standard MIDAS procedures.

The observations at the OAC were carried out from 2004 to 2006 with the F{\sc resco} \'echelle spectrograph 
mounted on the 91-cm telescope. The spectral resolution, as deduced from the FWHM of the lines of the 
Th-Ar calibration lamp, is $R\simeq 21\,000$. The 300-line/mm cross-disperser allowed to record about 
2500\,\AA ~at a time. The 19~\'echelle orders recorded by the CCD cover the 4310--6840\,\AA ~spectral 
region.We obtained 13, 6, and 12 spectra for RasTyc\,0524+6739, RasTyc\,1828+3506, and 
RasTyc\,2034+8253, respectively, with a S/N ranging from 30 to 100, depending on the star magnitude and 
sky conditions. In any case, giving the wide spectral range, the S/N was always adequate to perform good 
radial velocity (RV) measurements. The reduction of all these spectra was performed using the {\sc \'Echelle} 
task of IRAF\footnote{IRAF is distributed by the National Optical Astronomy Observatories, which are 
operated by the Association of Universities for Research in Astronomy, Inc., under cooperative agreement 
with the National Science Foundation.} package. \\

\onlfig{2}{
\begin{figure}[!ht] 
	\centering
	\includegraphics[width=4.0cm]{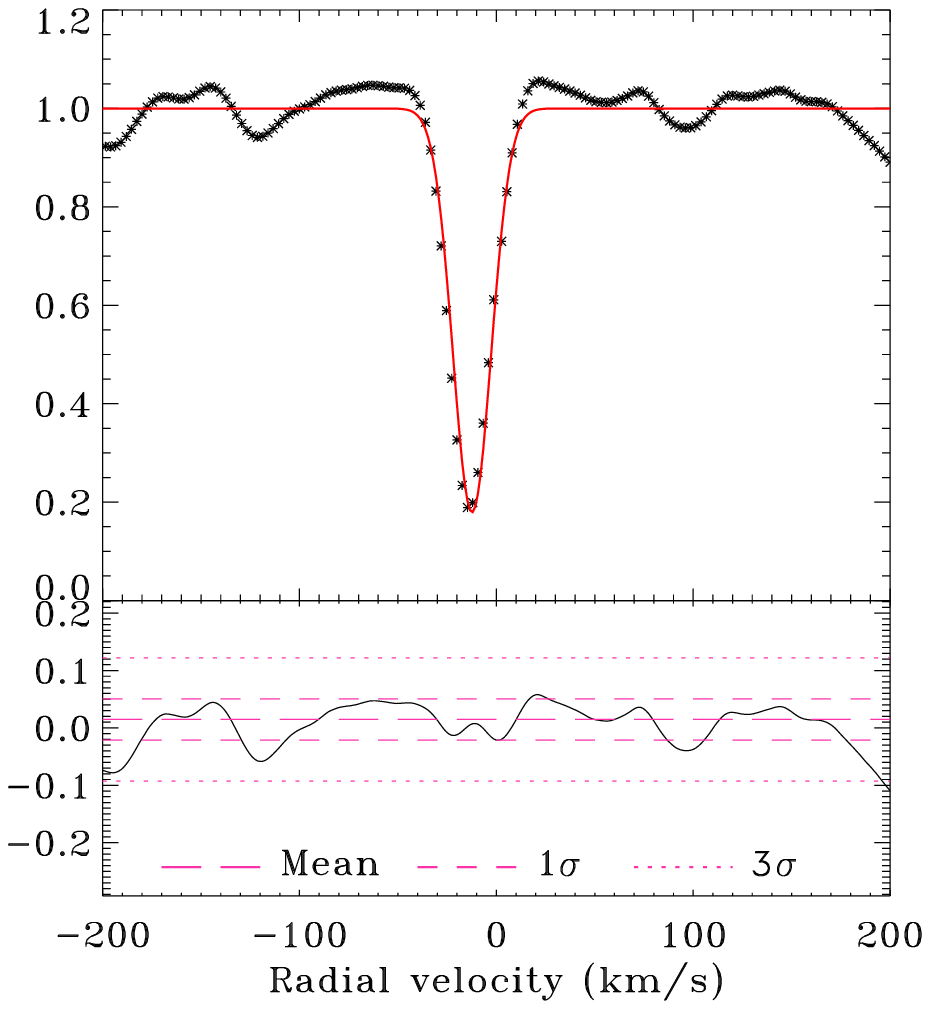}
	\includegraphics[width=4.0cm]{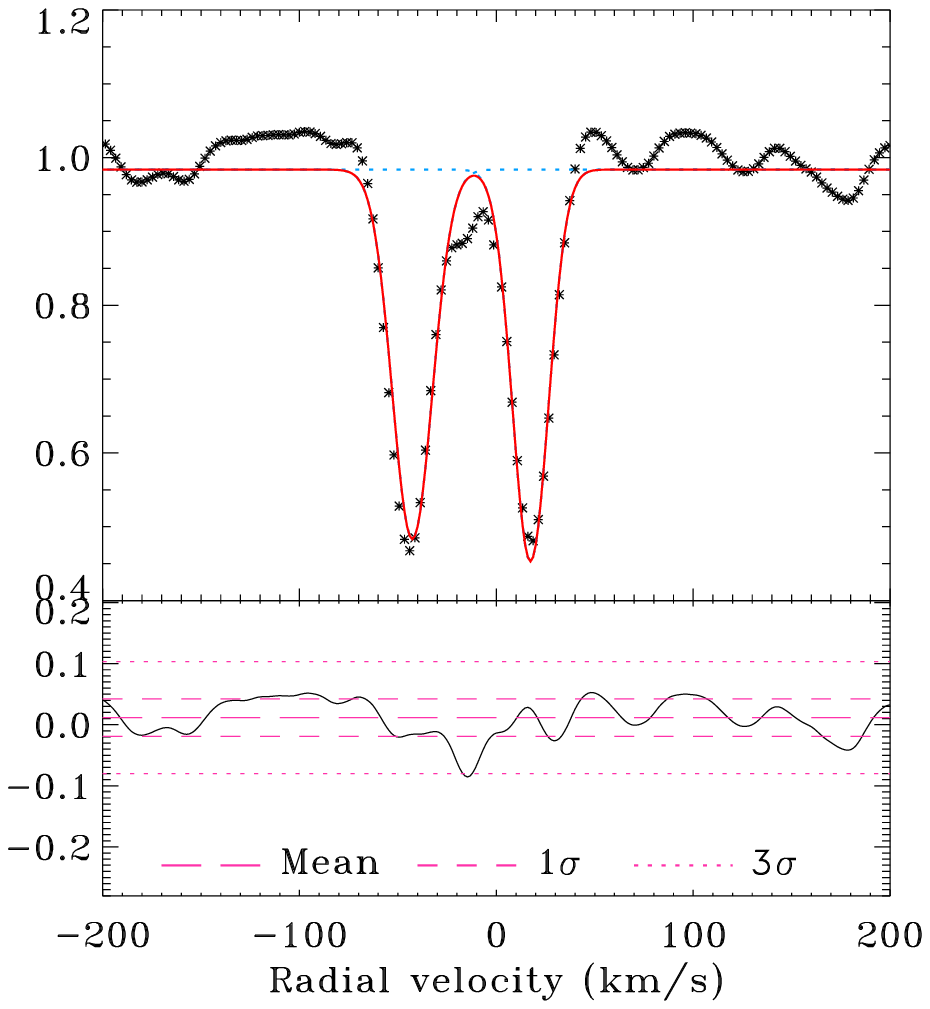}
	\includegraphics[width=4.0cm]{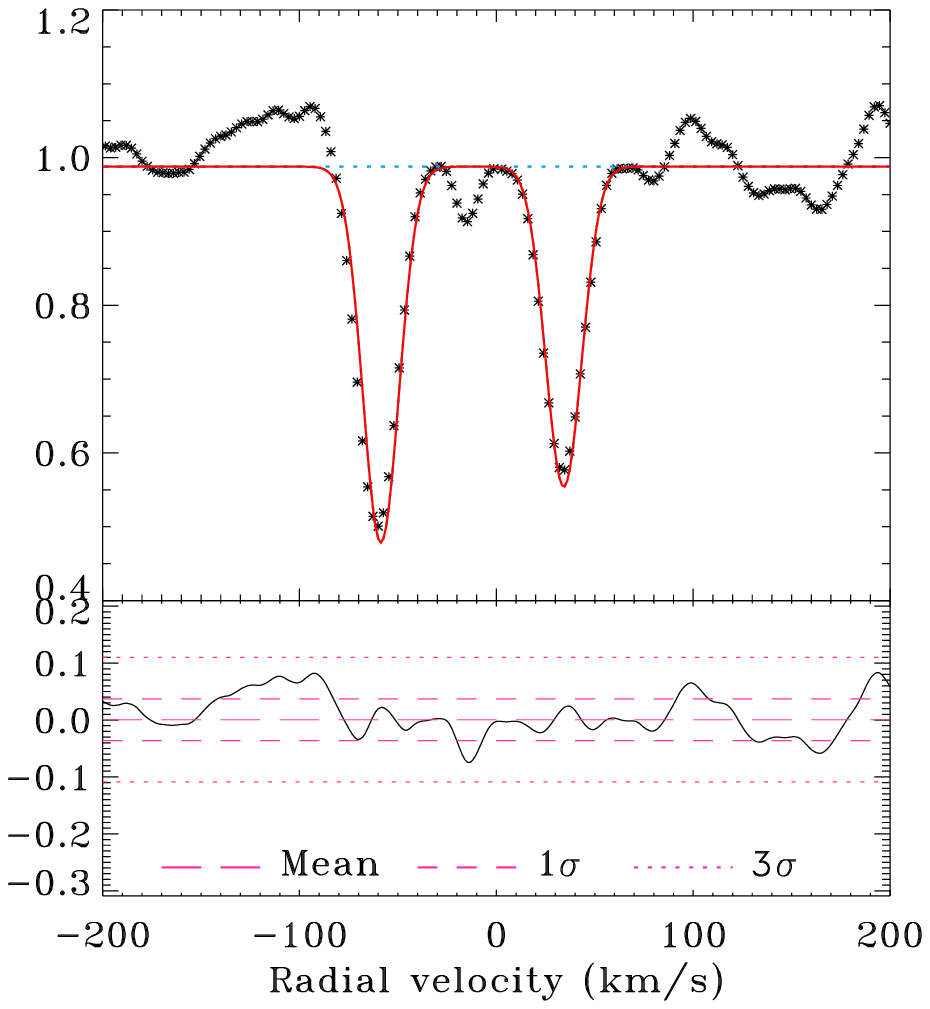}
	\includegraphics[width=4.0cm]{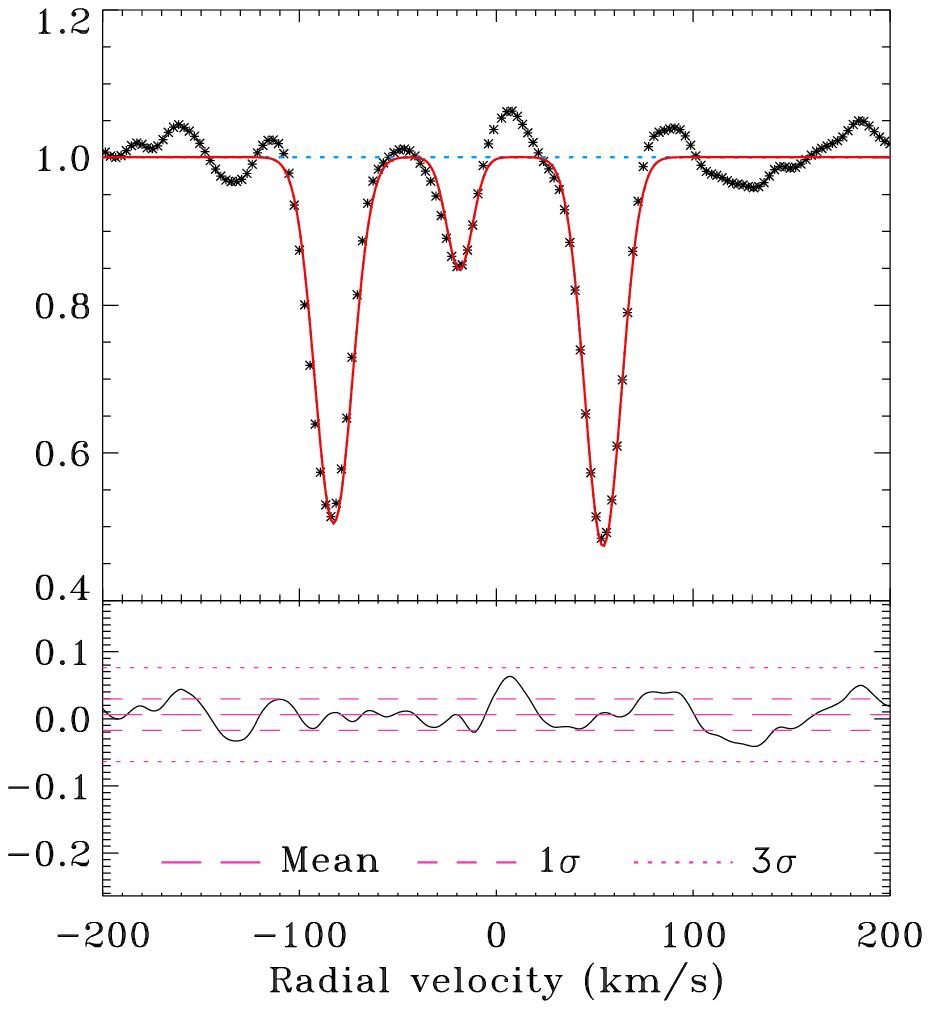}   
	\caption{\label{fig:CCF0524} The CCFs of RasTyc\,0524+6739 permit to emphasize the change of configuration 
	from a one-peak shape at a conjunction (left top panel) to a three-peak shape (right lower panel), passing 
	through phases of partial blending in which only two peaks are easily distinguishable (left lower panel 
	and right top panel). The RV uncertainty is about 1\,km\,s$^{-1}$ at our spectral resolution. On each panel, the 
	multiple Gaussian fits of the CCFs are over-plotted with full lines (top box) and the residuals of the 
	fits are plotted in the lower box.} 
\end{figure}
}

Photometric observations were carried out from 2004 to 2006 in the standard $UBV$ 
system with the 91-cm telescope of OAC. 
For each field of the three \textsl{RasTyc} sources studied here, we chose two or three stars 
with known $UBV$ magnitudes to be used as local standards for determining the photometric 
instrumental ``zero points''. Additionally, several standard stars, selected from the list of 
\citet{Lan92}, were also observed during the run in order to determine the transformation 
coefficients to the Johnson standard system. 
The data were reduced by means of the photometric data 
reduction package PHOT designed for the photoelectric photometry of OAC \citep{LoPr93}. 
The errors are typically $\sigma_V=0.006$, $\sigma_{B-V}=0.008$, and $\sigma_{U-B}=0.010$.

\subsection{Radial and projected rotational velocities determination}
\label{subsec:CCA}

From the analysis of the CCF (Figs.~\ref{fig:CCF0524},~\ref{fig:CCF1828}, and~\ref{fig:CCF2034})$^1$, 
we could accurately measure the RV and the projected rotational velocity ($v\sin i$) of each one of the stellar 
components. The RV measurements of these new systems are listed in Tables~$2$\,--\,$4$$^1$ together 
with their standard errors. 

\onlfig{3}{
\begin{figure}[!ht] 
	\centering
	\includegraphics[width=4.0cm]{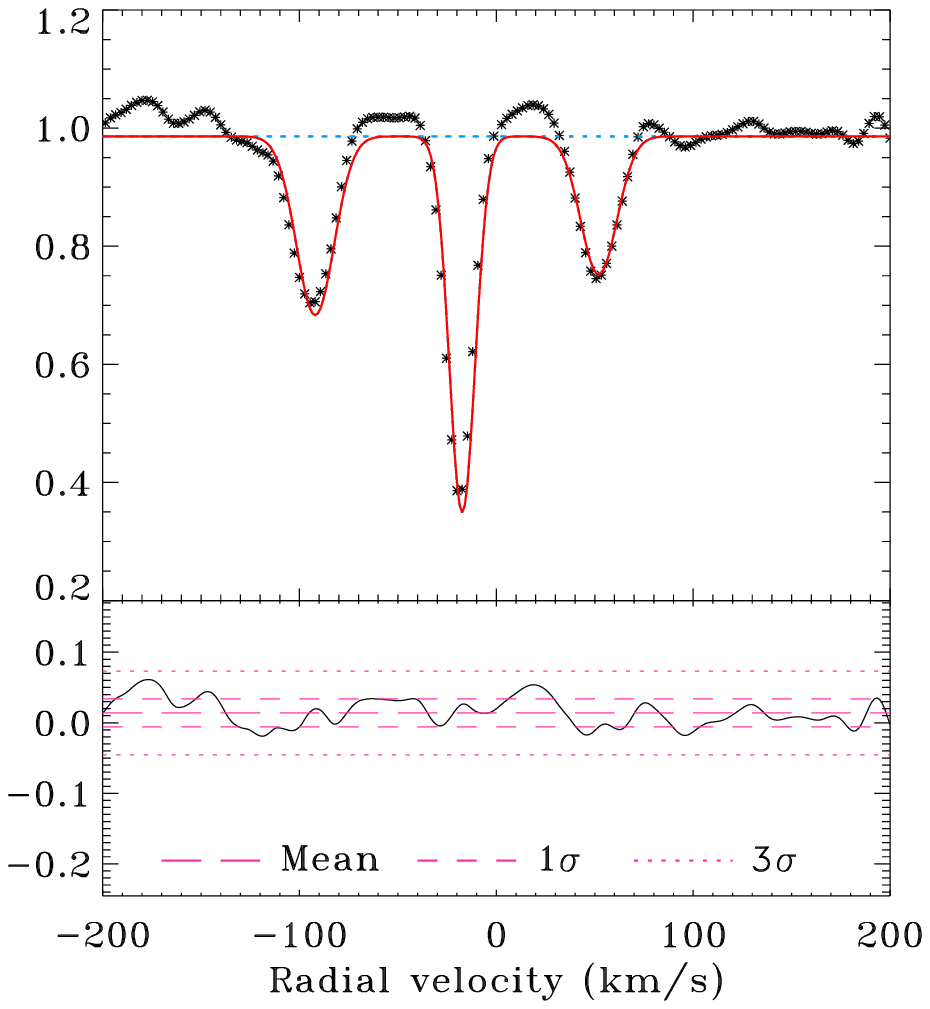}
	\includegraphics[width=4.0cm]{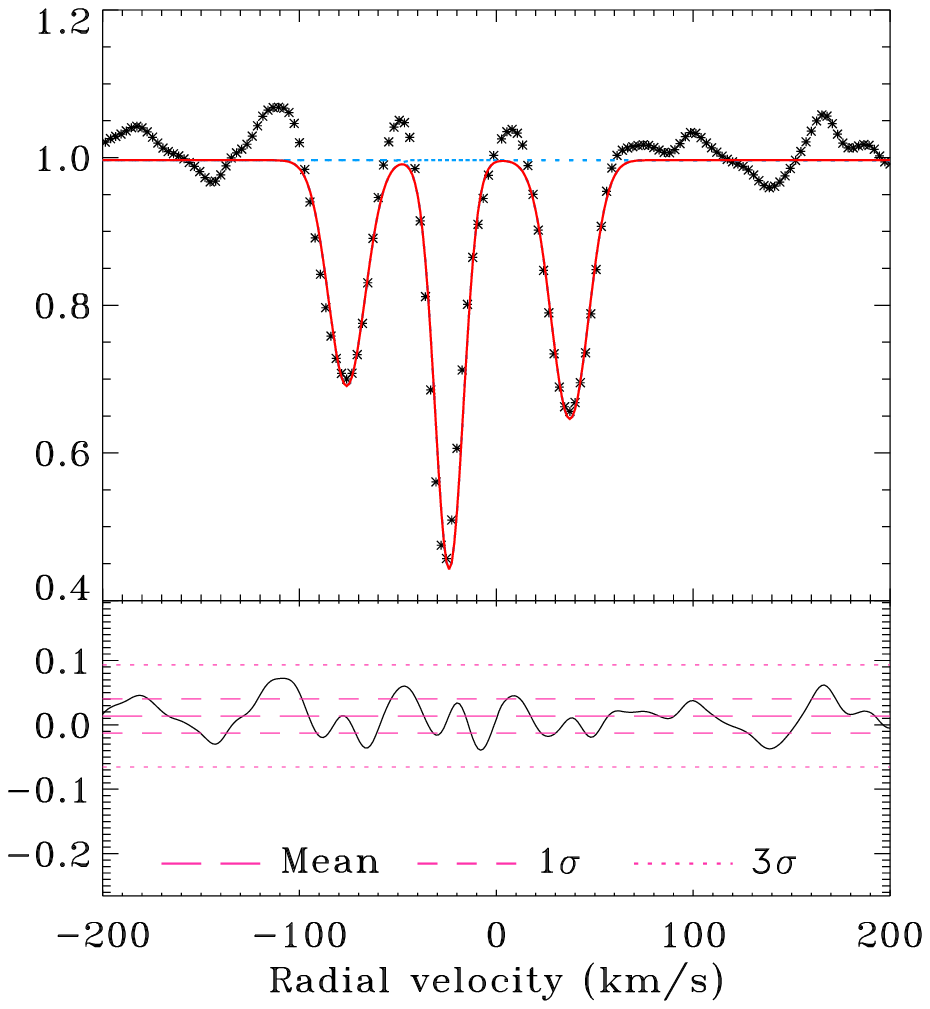}
	\caption{\label{fig:CCF1828} The CCFs of RasTyc\,1828+3506 clearly show three peaks with different widths. 
	The two components of the inner binary appear to rotate faster than the third star, probably due to spin-orbit synchronization.} 
\end{figure}
}

\onlfig{4}{
\begin{figure}[!ht] 
	\centering
	\includegraphics[width=4.0cm]{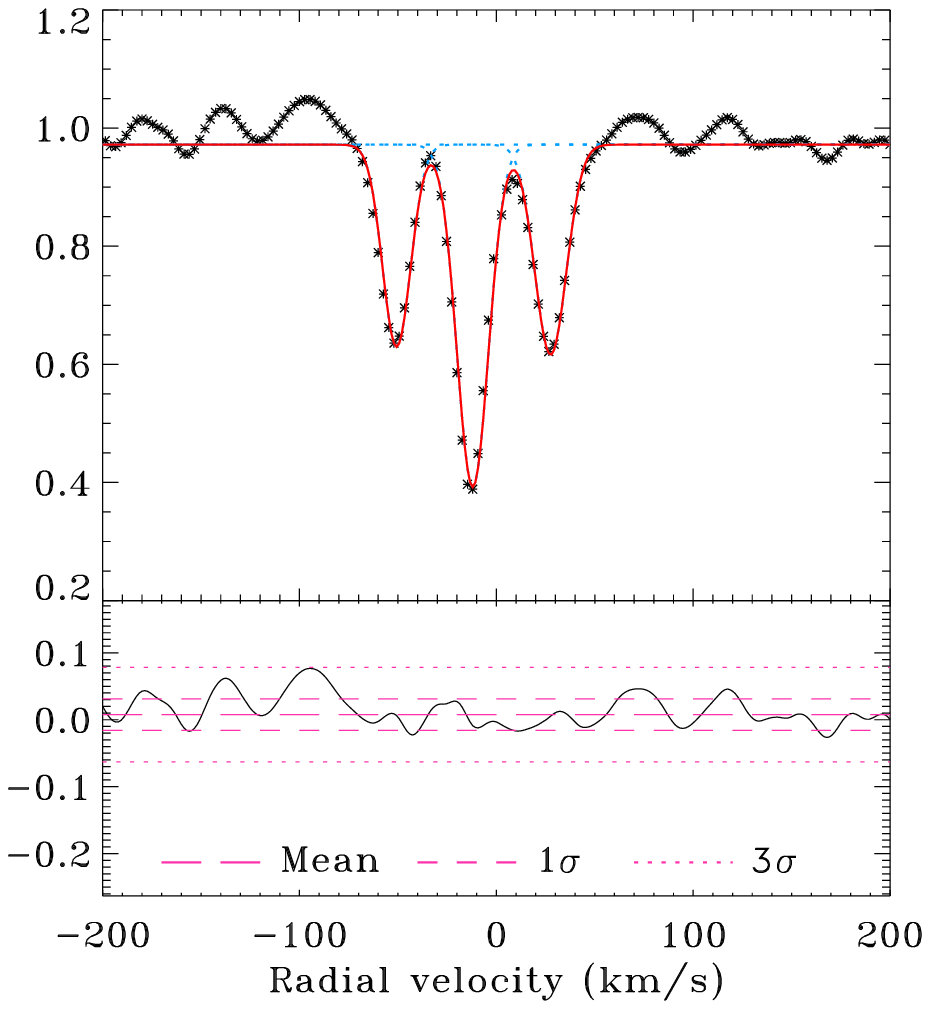}
	\includegraphics[width=4.0cm]{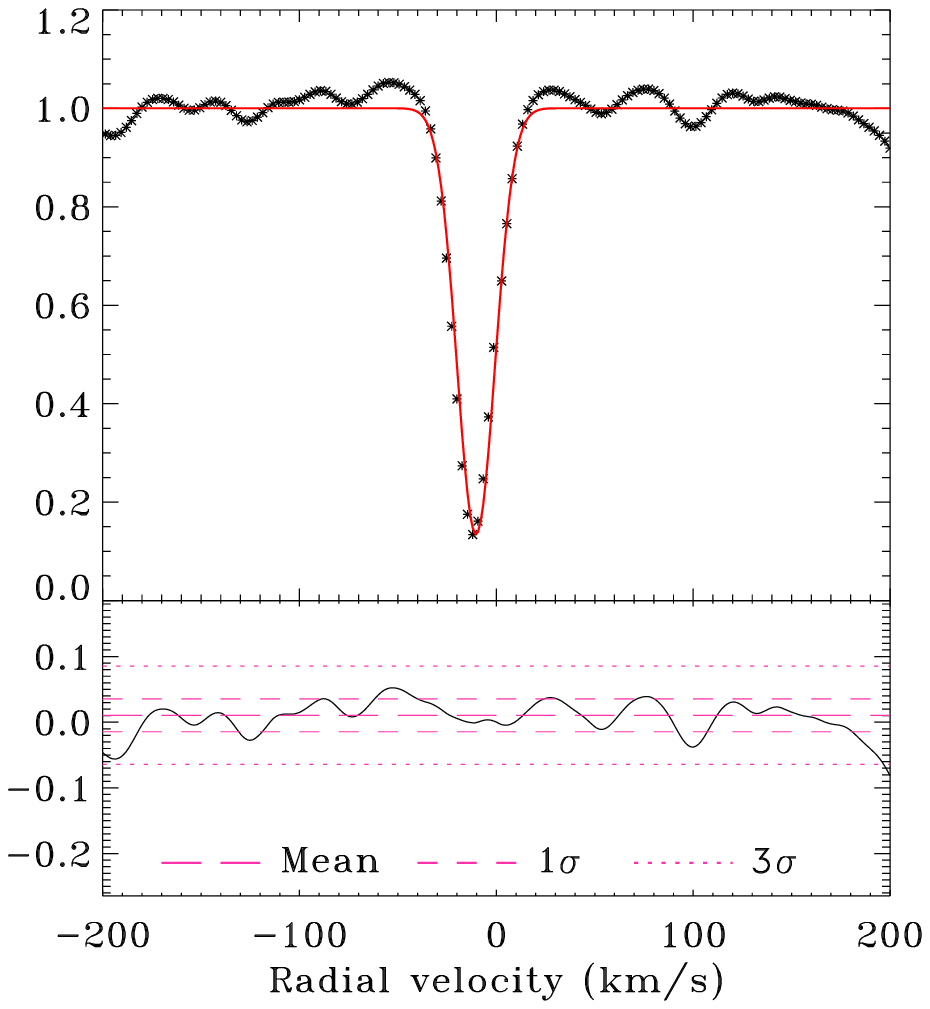}
	\caption{\label{fig:CCF2034} The CCFs of RasTyc\,2034+8253 at two orbital phases. The triple system 
	nature appears from the CCF (left panel), while it is completely hidden at another phase of observation 
	(right panel). }
\end{figure}
}

\onltab{2}{
\begin{table*}[!ht]
\caption[ ]{Radial velocity of the primary (more massive, $v_{\rm P}$), secondary ($v_{\rm S}$), and 
tertiary ($v_{\rm T}$) components of RasTyc\,0524+6739  from Aurelie (OHP) and FRESCO (OAC) 
spectra. The orbital phase has been computed according to the ephemeris $HJD_{\rm inf.conj.} = 
2\,451\,997.7654 + 3.6587 \times E$, with zero phase corresponding to the inferior conjunction for the 
primary component.}
\begin{center}
\begin{tabular}{ccrrrrrrc}
\hline
\hline
\noalign{\medskip}
  H.J.D.$^{a}$  &  Phase  & $v_{\rm P}$~~  & $\Delta v_{\rm P}$ & $v_{\rm S}$~~  & $\Delta v_{\rm S}$ & $v_{\rm T}$~~  & 
  $\Delta v_{\rm T}$ & Obs$^{c}$  \\
 {\scriptsize (2450000+)} &   & \multicolumn{2}{c}{(km s$^{-1}$)}   & \multicolumn{2}{c}{(km s$^{-1}$)} & 
 \multicolumn{2}{c}{(km s$^{-1}$)} & \\
\noalign{\medskip}
\hline
\noalign{\medskip}
  2216.45508  &  0.773  &    67.22     &   1.47  &  -96.83  &  1.51  &  -14.69  &  1.50 &  OHP \\
  2217.51416  &  0.062  &   -47.21     &   0.96  &   19.18  &  0.76  &  -12.33  &  0.77 &  OHP \\
  3570.57544  &  0.882  &    33.45     &   1.35  &  -71.51  &  1.58  &  -14.50  &  1.30 &  OHP \\
  3571.60986  &  0.165  &   -82.68     &   1.18  &   54.48  &  1.17  &  -12.72  &  1.37 &  OHP \\
  3572.51733  &  0.413  &   -59.27     &   1.30  &   33.95  &  1.30  &  -14.78  &  1.40 &  OHP \\
  3579.60791  &  0.351  &   -81.26     &   1.12  &   57.18  &  1.11  &  -12.64  &  1.12 &  OHP \\
  3580.57983  &  0.617  &    41.87     &   1.47  &  -73.43  &  1.46  &  -13.26  &  1.20 &  OHP \\
  3581.57983  &  0.890  &    35.23     &   1.46  &  -62.89  &  1.50  &  -14.08  &  1.20 &  OHP \\
  3582.57056  &  0.161  &   -78.56     &   1.52  &   55.35  &  1.52  &  -12.17  &  1.70 &  OHP \\
  3583.60596  &  0.444  &   -42.14     &   1.46  &   17.90  &  1.47  &  -14.14  &  1.40 &  OHP \\
  3585.61300  &  0.992  &  ---$^{b}$   &    ---  &     ---  &  ---   &  -12.60  &  1.24 &  OHP \\
  3586.60889  &  0.264  &   -95.28     &   1.31  &   68.24  &  1.42  &  -12.29  &  1.30 &  OHP \\
  3587.61035  &  0.538  &     7.17     &   1.50  &  -32.99  &  1.53  &    ---	&  ---  &  OHP \\  
  3588.58545  &  0.805  &    56.72     &   1.52  &  -91.28  &  1.51  &  -14.80  &  1.60 &  OHP \\
  3589.60693  &  0.084  &   -53.96     &   1.42  &   27.28  &  1.51  &  -13.05  &  0.90 &  OHP \\
  3590.61060  &  0.358  &   -78.68     &   1.47  &   52.60  &  1.52  &  -14.00  &  1.70 &  OHP \\ 
\noalign{\medskip} 
  3782.34390  &  0.763  &    67.87     &   0.96  &  -99.81  &  0.98  &  -15.74  &  2.28 &  OAC \\
  3791.33690  &  0.221  &   -90.43     &   0.82  &   69.26  &  0.90  &  -13.63  &  1.53 &  OAC \\
  3792.44580  &  0.524  &  ---$^{b}$   &    ---  &   ---    &  ---   &  -14.02  &  0.61 &  OAC \\
  3798.41420  &  0.155  &   -78.78     &   0.84  &   52.93  &  0.85  &  -14.10  &  1.85 &  OAC \\
  3799.38410  &  0.420  &   -56.04     &   0.88  &   29.01  &  0.89  &  -13.11  &  1.77 &  OAC \\
  3800.33420  &  0.680  &    69.56     &   2.89  &  -90.69  &  2.64  &  -13.54  &  4.16 &  OAC \\
  3809.33300  &  0.140  &   -72.26     &   1.01  &   49.68  &  0.96  &  -13.97  &  1.58 &  OAC \\
  3811.39950  &  0.705  &    68.02     &   1.02  &  -99.04  &  0.98  &  -16.24  &  2.47 &  OAC \\
  3812.38150  &  0.973  &  ---$^{b}$   &    ---  &   ---    &  ---   &  -14.13  &  0.46 &  OAC \\
  3826.29570  &  0.776  &    67.96     &   0.60  &  -95.73  &  0.64  &  -14.50  &  0.92 &  OAC \\
  3828.30120  &  0.324  &   -86.83     &   0.89  &   64.96  &  0.87  &  -13.16  &  1.87 &  OAC \\
  3833.30580  &  0.692  &    70.25     &   0.77  &  -92.96  &  0.88  &  -12.26  &  1.75 &  OAC \\
  3835.30440  &  0.238  &   -94.74     &   0.96  &   68.45  &  0.89  &  -16.45  &  1.91 &  OAC \\
\noalign{\medskip}
\hline	     
\noalign{\medskip}
\end{tabular}
\end{center}  
$^{a}$ Heliocentric Julian date at mid exposure.\\
$^{b}$ Blended CCF peaks.\\
$^{c}$ OHP = Observatoire de Haute Provence; OAC = Osservatorio Astrofisico di Catania.
\label{Tab:RV1}
\end{table*}
}

\onltab{3}{
\begin{table*}[!ht]
\caption[ ]{Radial velocity of the primary (more massive, $v_{\rm P}$), secondary ($v_{\rm S}$), and tertiary ($v_{\rm T}$) 
components of RasTyc\,1828+3506  from Aurelie (OHP) and FRESCO (OAC) spectra. The orbital phase has been 
computed only for the data of 2005 according to the ephemeris $HJD_{\rm inf.conj.} = 2\,452\,001.408 + 7.595 
\times E$, with zero phase corresponding to the inferior conjunction for the primary component.}
\begin{center}
\begin{tabular}{ccrrrrrrc}
\hline
\hline
\noalign{\medskip}
  H.J.D.$^{a}$  &  Phase  & $v_{\rm P}$~~  & $\Delta v_{\rm P}$ & $v_{\rm S}$~~  & $\Delta v_{\rm S}$ & $v_{\rm T}$~~  & 
  $\Delta v_{\rm T}$ & Obs$^{c}$  \\
 {\scriptsize (2450000+)} &   & \multicolumn{2}{c}{(km s$^{-1}$)}   & \multicolumn{2}{c}{(km s$^{-1}$)} & 
 \multicolumn{2}{c}{(km s$^{-1}$)} & \\
\noalign{\medskip}
\hline
\noalign{\medskip}
  3261.35425  &   ---   &  -75.11     &  1.17   &   37.34   &  1.27  &  -24.06  &  1.26 &  OHP \\
\noalign{\medskip} 
  3571.41016  &  0.715  &   47.23     &  1.33	&  -95.86   &  1.36  &  -19.98  &  1.25 &  OHP \\
  3572.36621  &  0.841  &   42.22     &  1.35	&  -79.96   &  1.16  &  -18.02  &  1.34 &  OHP \\
  3573.37378  &  0.974  &  ---$^{b}$  &  ---	&   ---     &  ---   &  -19.56  &  1.39 &  OHP \\
  3574.40454  &  0.109  &  -59.63     &  1.34	&   21.01   &  1.83  &  -19.32  &  1.20 &  OHP \\
  3575.42236  &  0.244  &  -93.37     &  1.15	&   53.36   &  1.15  &  -18.03  &  1.30 &  OHP \\
  3576.43481  &  0.377  &  -70.82     &  1.71	&   34.91   &  1.34  &  -17.80  &  1.27 &  OHP \\
  3577.36768  &  0.500  &  ---$^{b}$  &  ---	&   ---     &  ---   &  -19.55  &  1.40 &  OHP \\
  3578.36035  &  0.630  &   30.30     &  1.70	&  -67.73   &  2.82  &  -17.81  &  1.27 &  OHP \\
  3579.36890  &  0.763  &   51.28     &  1.33	&  -99.08   &  1.18  &  -17.38  &  1.26 &  OHP \\
  3580.35083  &  0.892  &   21.90     &  1.28	&  -65.14   &  1.34  &  -17.90  &  1.26 &  OHP \\
  3581.38745  &  0.029  &  ---$^{b}$  &  ---	&   ---     &  ---   &  -18.81  &  1.39 &  OHP \\
  3582.36328  &  0.157  &  -83.31     &  1.60	&   32.82   &  1.22  &  -17.96  &  1.20 &  OHP \\
  3583.37573  &  0.291  &  -92.79     &  1.34	&   51.14   &  1.36  &  -17.43  &  1.26 &  OHP \\  
\noalign{\medskip} 
  3648.39840  &  0.852  &   40.56     &  1.23   &  -82.21   &  1.04  &  -15.66  &  0.69 &  OAC \\
  3657.36380  &  0.032  &  -43.51     &  1.24   &   ---     &  ---   &  -13.00  &  0.71 &  OAC \\ 
\noalign{\medskip}   
  3927.49000  &   ---   &  -66.97     &  1.64	& ---$^{b}$ &  ---   &    ---   &  ---  &  OAC \\
  3931.45260  &   ---   &   30.57     &  5.23	&  -82.14   &  2.13  &   -8.81  &  2.32 &  OAC \\ 
  3932.46490  &   ---   &   44.26     &  2.37	&  -96.87   &  2.68  &   -7.45  &  2.22 &  OAC \\
  3940.42350  &   ---   &   38.43     &  1.62	&  -98.48   &  1.11  &   -8.97  &  1.07 &  OAC \\ 
\noalign{\medskip}
\hline	     
\noalign{\medskip}
\end{tabular}
\end{center}  
$^{a}$ Heliocentric Julian date at mid exposure.\\
$^{b}$ Blended CCF peaks.\\
$^{c}$ OHP = Observatoire de Haute Provence; OAC = Osservatorio Astrofisico di Catania.
\label{Tab:RV2}
\end{table*}
}

\onltab{4}{
\begin{table*}[!ht]
\caption[ ]{Radial velocity of the primary (more massive, $v_{\rm P}$), secondary ($v_{\rm S}$), and tertiary ($v_{\rm T}$) 
components of RasTyc\,2034+8253  from Aurelie (OHP) and FRESCO (OAC) spectra. The orbital phase has been 
computed according to the ephemeris $HJD_{\rm inf.conj.} = 2\,452\,010.7086 + 4.9538 \times E$, with zero phase 
corresponding to the inferior conjunction for the primary component.}
\begin{center}
\begin{tabular}{ccrrrrrrc}
\hline
\hline
\noalign{\medskip}
  H.J.D.$^{a}$  &  Phase  & $v_{\rm P}$~~  & $\Delta v_{\rm P}$ & $v_{\rm S}$~~  & $\Delta v_{\rm S}$ & $v_{\rm T}$~~  & 
  $\Delta v_{\rm T}$ & Obs$^{c}$  \\
 {\scriptsize (2450000+)} &   & \multicolumn{2}{c}{(km s$^{-1}$)}   & \multicolumn{2}{c}{(km s$^{-1}$)} & 
 \multicolumn{2}{c}{(km s$^{-1}$)} & \\
\noalign{\medskip}
\hline
\noalign{\medskip}
  2475.51587  &  0.828  &   20.57      &   1.27  &  -43.24  &  1.41  &  -10.96  &  1.26  &  OHP \\
  2482.39160  &  0.216  &  -50.12      &   1.42  &   29.29  &  1.28  &   -9.99  &  1.39  &  OHP \\
  3213.54565  &  0.811  &   25.34      &   1.28  &  -47.15  &  1.36  &  -10.75  &  1.27  &  OHP \\
  3214.55176  &  0.014  &   ---$^{b}$  &    ---  &    ---   &  ---   &  -10.36  &  1.41  &  OHP \\
  3215.54883  &  0.215  &  -50.62      &   1.28  &   27.43  &  1.41  &  -11.96  &  1.26  &  OHP \\
  3216.51929  &  0.411  &  -31.46      &   1.28  &    5.58  &  1.44  &  -13.23  &  1.41  &  OHP \\
  3261.42578  &  0.476  &   ---$^{b}$  &    ---  &    ---   &  ---   &  -10.35  &  1.31  &  OHP \\
  3261.59082  &  0.510  &   ---$^{b}$  &    ---  &    ---   &  ---   &  -10.71  &  1.37  &  OHP \\
  3262.40967  &  0.675  &   23.25      &   1.42  &  -45.85  &  1.42  &  -10.79  &  1.27  &  OHP \\
  3264.55518  &  0.108  &  -35.53      &   1.36  &   13.24  &  1.43  &  -10.83  &  1.28  &  OHP \\
  3265.35815  &  0.270  &  -49.28      &   1.28  &   29.36  &  1.28  &   -9.46  &  1.26  &  OHP \\
  3265.60254  &  0.319  &  -45.91      &   1.42  &   24.57  &  1.42  &  -10.71  &  1.27  &  OHP \\
  3266.34009  &  0.468  &   ---$^{b}$  &    ---  &    ---   &  ---   &  -10.02  &  1.31  &  OHP \\
  3267.31812  &  0.666  &   23.25      &   1.27  &  -45.01  &  1.48  &   -9.68  &  1.26  &  OHP \\
  3268.33228  &  0.870  &   17.08      &   1.34  &  -39.55  &  1.43  &   -9.69  &  1.26  &  OHP \\
  3570.51855  &  0.871  &   17.60      &   1.27  &  -39.72  &  1.27  &  -10.00  &  1.28  &  OHP \\
  3571.46240  &  0.062  &   ---$^{b}$  &    ---  &    ---   &  ---   &  -10.33  &  1.29  &  OHP \\
  3572.46777  &  0.265  &  -50.64      &   1.28  &   28.86  &  1.40  &  -10.66  &  1.27  &  OHP \\
\noalign{\medskip} 
  3218.52790  &  0.817  &   24.88      &   1.38  &  -46.29  &  1.57  &   -9.65  &  0.79  &  OAC \\
  3219.52590  &  0.018  &   ---$^{b}$  &   ---   &    ---   &  ---   &  -10.71  &  0.24  &  OAC \\
  3220.53390  &  0.222  &  -49.63      &   1.07  &   27.41  &  1.07  &  -11.17  &  0.61  &  OAC \\
  3224.54270  &  0.031  &   ---$^{b}$  &   ---   &    ---   &  ---   &  -11.53  &  0.36  &  OAC \\
  3225.52580  &  0.229  &  -50.25      &   1.82  &   28.06  &  1.62  &  -11.20  &  0.59  &  OAC \\
  3256.57340  &  0.497  &   ---$^{b}$  &   ---   &    ---   &  ---   &   -9.25  &  0.36  &  OAC \\
  3273.38010  &  0.889  &   11.83      &   1.12  &  -38.13  &  1.32  &  -13.28  &  0.78  &  OAC \\
  3275.41070  &  0.299  &  -48.73     &    1.04	 &   26.07  &  1.14  &  -11.70  &  0.61  &  OAC \\
  3279.34730  &  0.094  &  -30.29      &   1.27	 &   13.36  &  1.44  &  -10.36  &  0.60  &  OAC \\
  3281.39960  &  0.508  &   ---$^{b}$  &   ---   &    ---   &  ---   &   -9.19  &  0.33  &  OAC \\
  3285.45010  &  0.326  &  -46.27      &   1.07  &   23.98  &  1.14  &  -11.57  &  0.62  &  OAC \\
  3354.34210  &  0.233  &  -48.64      &   1.05  &   28.44  &  1.13  &  -10.19  &  0.40  &  OAC \\
\noalign{\medskip} 
\hline	     
\noalign{\medskip}
\end{tabular}
\end{center}  
$^{a}$ Heliocentric Julian date at mid exposure.\\
$^{b}$ Blended CCF peaks.\\
$^{c}$ OHP = Observatoire de Haute-Provence; OAC = Osservatorio Astrofisico di Catania.
\label{Tab:RV3}
\end{table*}
}

\begin{table*}
\caption{Orbital parameters of the three systems. The errors on the last significant 
digit are enclosed in parenthesis. P = Primary and S = Secondary.}
\begin{center}
\begin{tabular}{lllcccccc}
\hline
\hline
\noalign{\medskip}
   Name   & HJD0 &  $P_{\rm orb}$  & $e$ & $\omega$    &   $\gamma$  &	 $k$	   &  $M\sin^3i$    & $M_P$/ $M_S$ \\
             & {\scriptsize(2\,450\,000+)} & (days) &  & ($\degr$) &  (km\,s$^{-1}$)  &  (km\,s$^{-1}$)  &  (M$_{\odot}$) &  \\
             &      &   	      &		&	&         &	[P/S]	    &   [P/S]	     &  \\
\noalign{\medskip}
\hline
\noalign{\medskip}
RasTyc\,0524+6739 & 2212.02(5)$^{\,\rm a}$ & 3.65884(2) & 0.042(3) & 294(5) & $-13.4(2)$ & 82.2(3)/84.5(3) & 0.889(6)/0.865(6) & 1.028(4)\\
RasTyc\,1828+3506 & 3565.95(4)$^{\,\rm b}$    & 7.595(5)   & 0        &  ---   & $-21.5(4)$ & 72.4(5)/73.7(5) & 1.24(2)/1.22(2)   & 1.018(9)\\
RasTyc\,2034+8253 & 2471.3(3)$^{\,\rm b}$  & 4.9543(2)  & 0        &  ---   & $-11.2(3)$ & 38.8(3)/39.5(3) & 0.124(2)/0.122(2) & 1.02(1)\\
\hline 
\noalign{\medskip}
\end{tabular}
\begin{tabular}{cc}
	$^{\rm a}$ Heliocentric Julian Date (HJD) of the periastron passage; & $^{\rm b}$ HJD of the inferior conjunction of the primary (more massive) component.\\
\end{tabular}
\end{center}
\label{Tab:ParamOrb} 
\end{table*}

 \begin{figure*}[ht]
  \begin{center}
  \includegraphics[width=6.0cm]{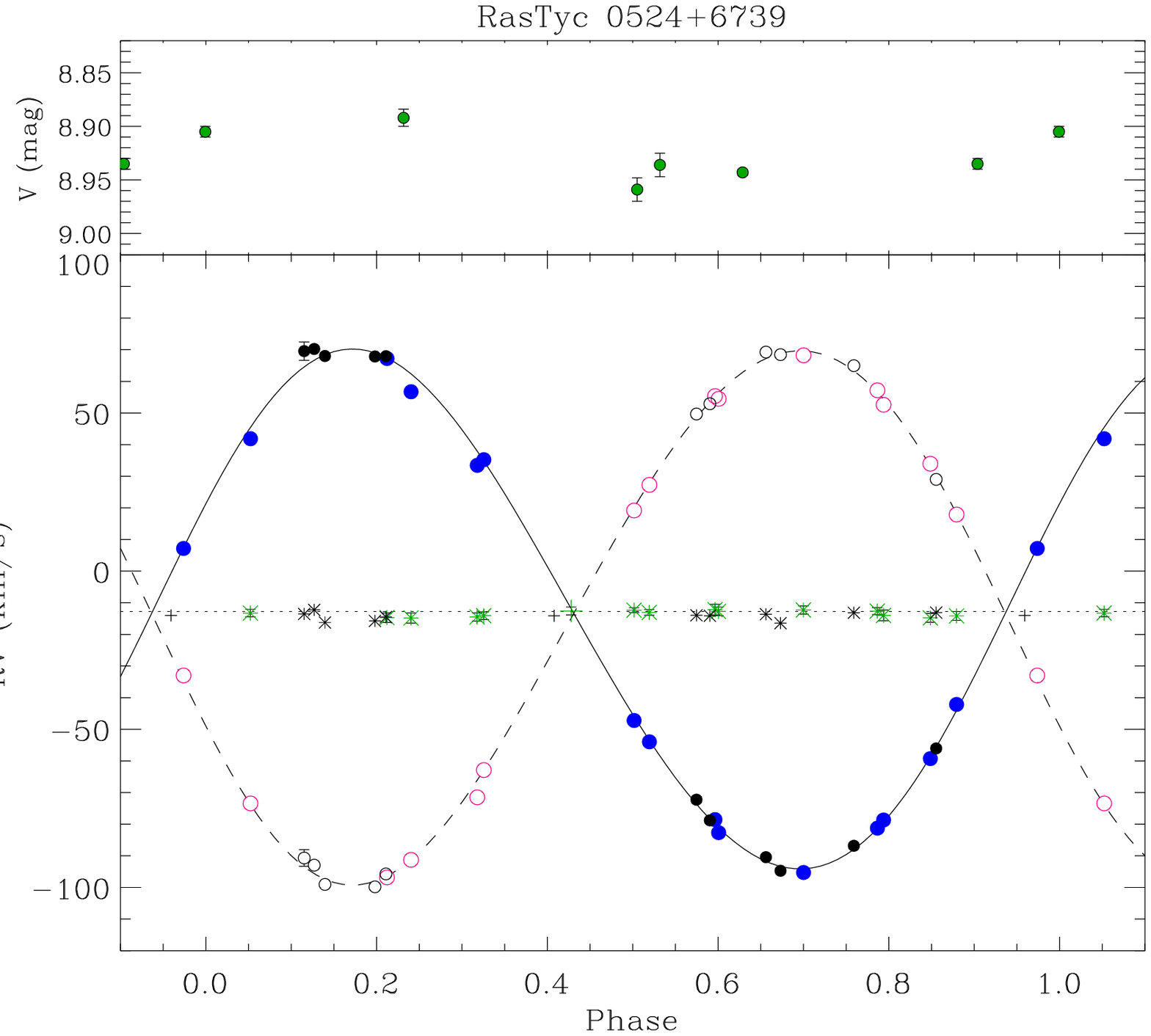}
  \includegraphics[width=6.0cm]{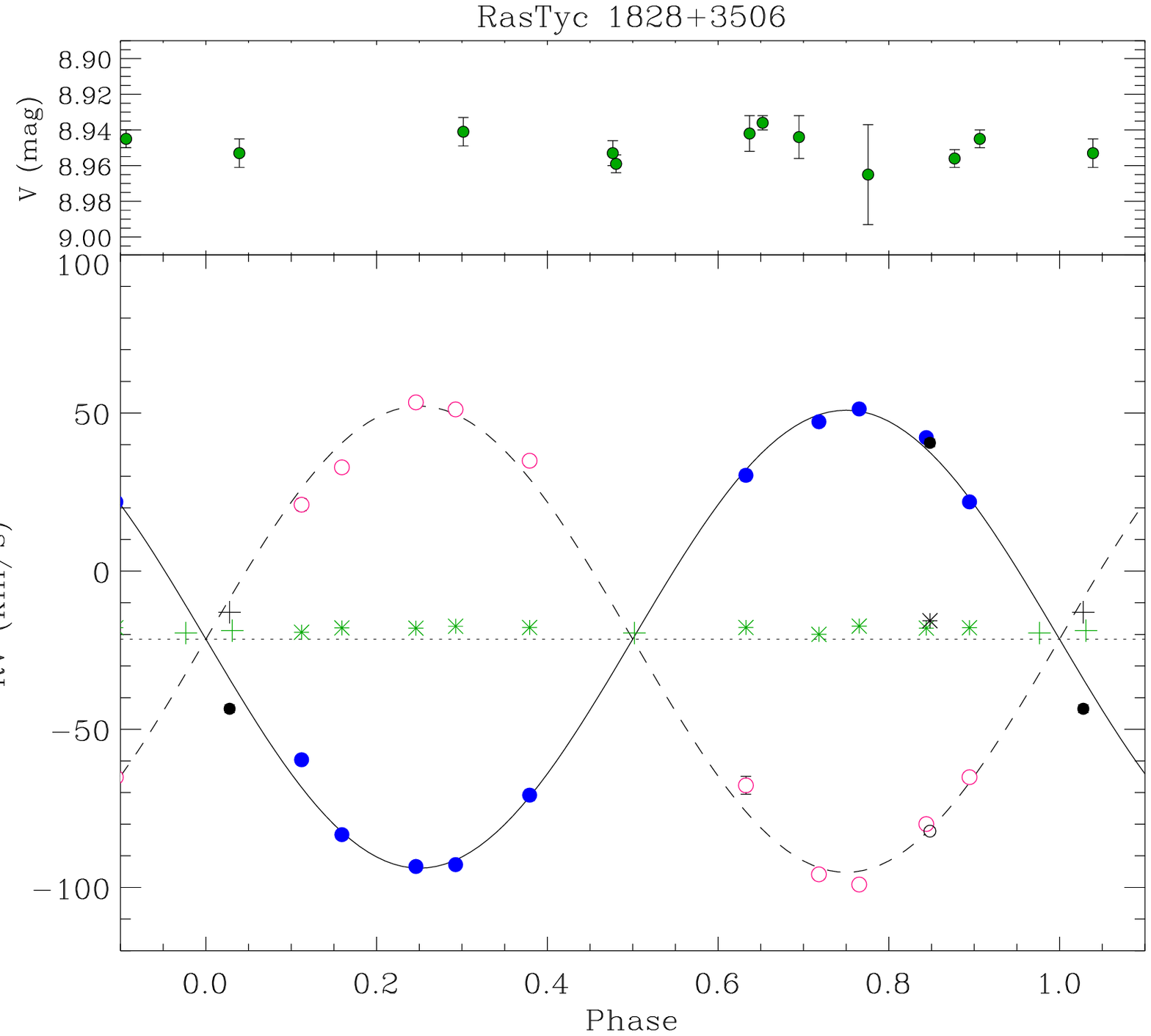}
  \includegraphics[width=6.0cm]{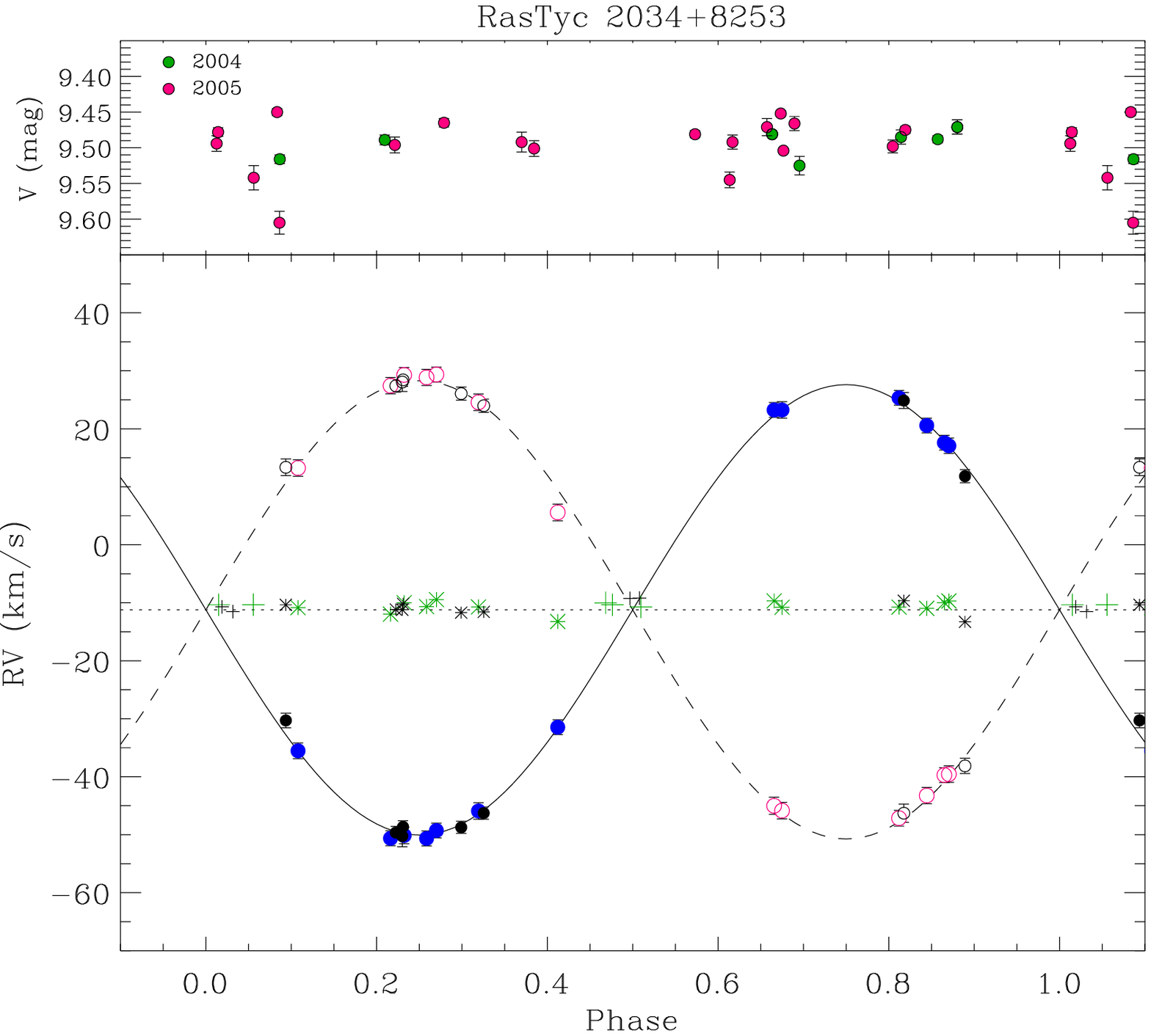}
  \end{center}
\caption{Radial velocity curves of the new three {\it RasTyc} triple systems. Large symbols refers to 
A{\sc urelie} data while smaller symbols are used for F{\sc resco} spectrograph data. Filled and 
open circles for the primary (more massive) and secondary component of the inner binaries have been 
used, respectively. In each panel, the solid and dashed lines represents the orbital solutions for the 
primary and secondary component, respectively, whereas the dotted line represents the barycenter of 
the inner binary. The asterisks are used for the tertiary component and plus symbols refer to blended 
RV values. The RV errors are always smaller than, or comparable to, the symbol size. The $V$ photometry 
is displayed, as a function of the orbital phase, on the top panel of each box. 
}
\label{fig:RV}
\end{figure*}

\subsection{Spectral types and stellar parameters}
\label{subsec:SpT3}

With the aim to have a first guess of the spectral types of these objects, we applied the ROTFIT code
\citep{Frasca03,Frasca2006} to spectra for which only one peak is visible in the CCF. 

For the evaluation of the spectral type of each individual component, we analyzed the spectra for which the 
CCF shows three distinct peaks. We used another IDL code, similar to ROTFIT, to estimate the spectral type, the astrophysical parameters (APs), and the continuum flux contribution of the tertiary component. The 
subtraction of the spectrum fitted to this component provided us with a ``cleaned" spectrum of the inner binary 
that we could analyze with COMPO2 \citep{Frasca2006} in the usual way. 
In Sect.~\ref{sec:TripleProp},  we discuss the results of this procedure concerning the spectral type and 
the contribution to the continuum flux (weight) as well as the values (the weighted mean of the best 100 
combinations) of the APs, namely the effective temperature ($T_{\rm eff}$), gravity ($\log g$), and metallicity 
([Fe/H]), for the three components of each triple system. 

Moreover, we applied the ``spectral subtraction'' technique \citep[e.g.,][]{FraCa94,Montes1995} to measure 
the \ion{Li}{i} equivalent width, $EW$(Li), of each stellar component (Fig.~\ref{fig:TripleLi}, bottom of each 
box) and used the 
\citet{PavMag96} calculations to deduce a Lithium abundance, $\log N$(Li). From these parameters, we 
estimated the age of these three systems in Sect.~\ref{sec:AGE}.

\section{Results and discussion}
\label{sec:Discussion}

\subsection{Orbital solutions}
\label{subsec:RadVel}

We initially searched for eccentric orbits and we found low eccentricity values ($e=0.042\pm0.003$, 
$e=0.026\pm0.007$, and $e=0.020\pm0.012$ for RasTyc\,0524+6739, RasTyc\,1828+3506, and 
RasTyc\,2034+8253, respectively). Following the precepts of \citet[][Eq.\,22]{Lucy71}, we considered 
as significant only the eccentricity of RasTyc\,0524+6739 and adopted $e=0$ (circular orbits) for the other 
two systems. The {\sc Curvefit} routine \citep{Bevi69} was used to fit the observed RV curves and  
to determine the orbital parameters and their standard errors for each inner binary (Table~\ref{Tab:ParamOrb}).
The observed RV curves of the three triple systems are displayed in Fig.~\ref{fig:RV}. 

For the two RasTyc systems with the shortest orbital periods, the RV value of the tertiary component is 
very close to the barycentric velocity ($\gamma$) of the inner binary during different observing seasons. 
Thus, we could not try any evaluation of the orbital period of the tertiary component. 
As the vast majority of the already known triple systems, each of our sources consists of a 
short-period inner binary with a third component orbiting around the close pair in a long-period orbit. These 
systems display a typical ``hierarchical'' configuration \citep{Evans68}. In particular, RasTyc\,2034+8253 was 
already known as a visual binary \citep{Mlr76}. 
From the observations of \citet{Mlr76, Mlr78, Mlr90} and \citet{Fabricius2002}, this component seems to have 
a regular evolution because the position angle and the separation are constantly growing. 
Thus, its orbital period must be significantly greater than the period of observations (about 20 years) 
implying a degree of hierarchy $X\gg1500$, where $X$ is defined as the ratio of the ``external'' period 
(orbit of tertiary component around the center of mass to the inner binary) to the ``internal'' period 
(that of the inner binary). 

\begin{figure}[ht]
  \begin{center}
  \includegraphics[width=9.0cm]{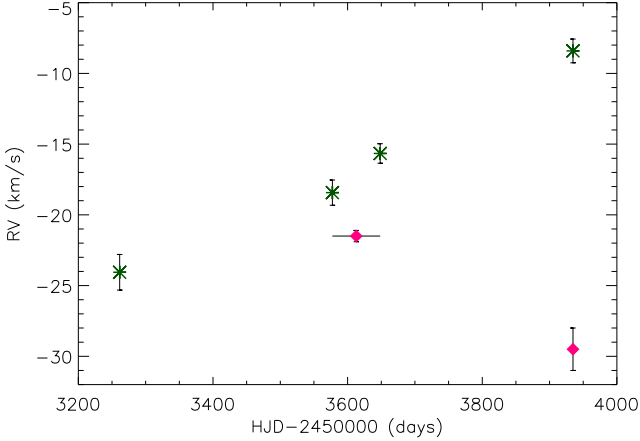}
  \end{center}
\caption{RV variation for the tertiary component of RasTyc\,1828+3506 (asterisks) during the four 
	seasons of observations which span from 2004 to 2006. The barycenter RV of the 
	inner binary in 2005 and 2006 is also displayed by two diamonds.}
\label{fig:rx1828_ter}
\end{figure}

On the contrary, in 2005, the tertiary component of RasTyc\,1828+3506 displays a RV 
systematically higher (3.3\,km\,s$^{-1}$) than that of the barycenter of the inner binary 
(see Fig.~\ref{fig:RV}, middle panel). Moreover, 
we found a highly significant RV increase (more than 15\,km\,s$^{-1}$) for the tertiary component 
during the four seasons of observations (Fig.~\ref{fig:rx1828_ter}).
This relevant RV variation prevented us from using all the RV values to obtain the orbital solution of the inner 
binary. Thus, we deduced the orbital parameters from the solution of OHP and OAC data acquired in 2005 only. The four RVs obtained at OAC in 2006 allowed us to estimate the barycentric velocity 
$\gamma=-29.5\pm1.5$\,km\,s$^{-1}$ at that epoch, by adopting the values of semi-major axes, 
$k_P=72.4$\,km\,s$^{-1}$ and $k_S=73.7$\,km\,s$^{-1}$, found from the solution of 2005 data. We could not 
evaluate the orbital period of the tertiary, but, from the data trend, we can argue that the period must be larger 
than 6 years (but presumably shorter than some tens years) implying $X\geq300$. New 
observations will enable us to obtain the solution for the tertiary orbit. 

Using the empirical stability criterion described by \citet{OP2000}, we can conclude that these three newly 
discovered triple systems are in a gravitationally stable orbital configuration. 

\subsection{Photometric variability}
\label{sec:Phot}

The photometric monitoring of our sources in $UBV$ bands allowed us to determine the mean 
magnitudes and color indices as well as to detect any eventual variability. 
The average values of $V$, $B-V$, and $U-B$ are reported in Table~\ref{Tab:PhotoData}.

RasTyc\,0524+6739, the system with the shortest orbital period, is the only one showing a clear 
variation that is likely correlated with the orbital period. Unfortunately, because of poor weather 
conditions during the winter observing season, the photometric data were obtained only on six 
nights and are not sufficient to cover evenly the light curve. However, we estimated the amplitude 
of the $V$ light curve to about 0$\fm$07.

For RasTyc\,1828+3506 and RasTyc\,2034+8253 there is no indication of rotational modulation in the
$V$ magnitude, in spite of the larger number of data. However, some indication of a stochastic variation 
comes out for RasTyc\,2034+8253, whose magnitude varies in the range 9$\fm$45\,--\,9$\fm$60 with errors 
lower than 0$\fm$02. No clear periodicity was found in these data. 

The detection of a $V$ modulation only on the system with the shortest orbital period 
($P_{\rm orb}\simeq$\,3.6 days) is in line with the well established enhancement of magnetic
activity in the components of close binaries due both to fast rotation and nearness. 
Moreover, tidal interaction in these binaries leads to synchronization between orbital and rotational periods of 
both components. 
It is worth noticing that the periods of all these three systems are smaller than, or very close to, the cut-off 
value of 7.56 days found by \citet{Melo2001} for orbital circularization in Pre-Main Sequence (PMS) binaries. 
According to \citet{ZB89}, for close late-type binaries with masses ranging from 0.5 to 1.25\,M$_{\sun}$, the 
cut-off period may be as long as 7.2 to 8.5 days, depending on the masses and on the assumptions of the 
initial conditions. So, being all these systems older than PMS stars, they should be already circularized, in 
substantial agreement with the results from the solution of their RV curves. The non-detection 
of photometric variation in RasTyc\,1828+3506 could be related both to the fairly long orbital/rotational 
period (7.595 days) and to the relatively early spectral types of the system components with shallower 
convective envelopes and, consequently, with a reduced dynamo action compared to cooler stars with 
the same rotation rate.

\begin{table}[t]
\caption{Photometric data and distance determination for the
three systems. The error on the last significant digit is enclosed in parenthesis.}
\begin{center}
\begin{tabular}{llllc}
\hline
\hline
\noalign{\medskip}
Name & ~~~~~$V$ & ~~$B-V$	   & ~~$U-B$ &~~~~Dist.\\
        & ~~(mag)       & ~~(mag)      & ~~(mag) & ~~(pc) \\
\noalign{\medskip}
\hline
\noalign{\medskip}
RasTyc\,0524+6739 & 8.892(7)$^{\,\rm a}$ 	& 0.773(7) & 0.235(9) & ~~75 $\pm$ 20 \\
RasTyc\,1828+3506 & 8.951(8)             		& 0.578(6) & 0.015(8) &   115 $\pm$ 25 \\
RasTyc\,2034+8253 & 9.49(2)              		& 0.96(1)   & 0.70(3)    & ~~80 $\pm$ 20 \\
\hline
\noalign{\medskip}
\end{tabular}
\begin{tabular}{l}
$^{\rm a} V$ magnitude at maximum brightness.\\
\end{tabular}
\end{center}
\label{Tab:PhotoData}
\end{table}

\subsection{Astrophysical parameters and other properties}
\label{sec:TripleProp}

\begin{figure}[!t] 
  \begin{center}
	\includegraphics[width=8.5cm]{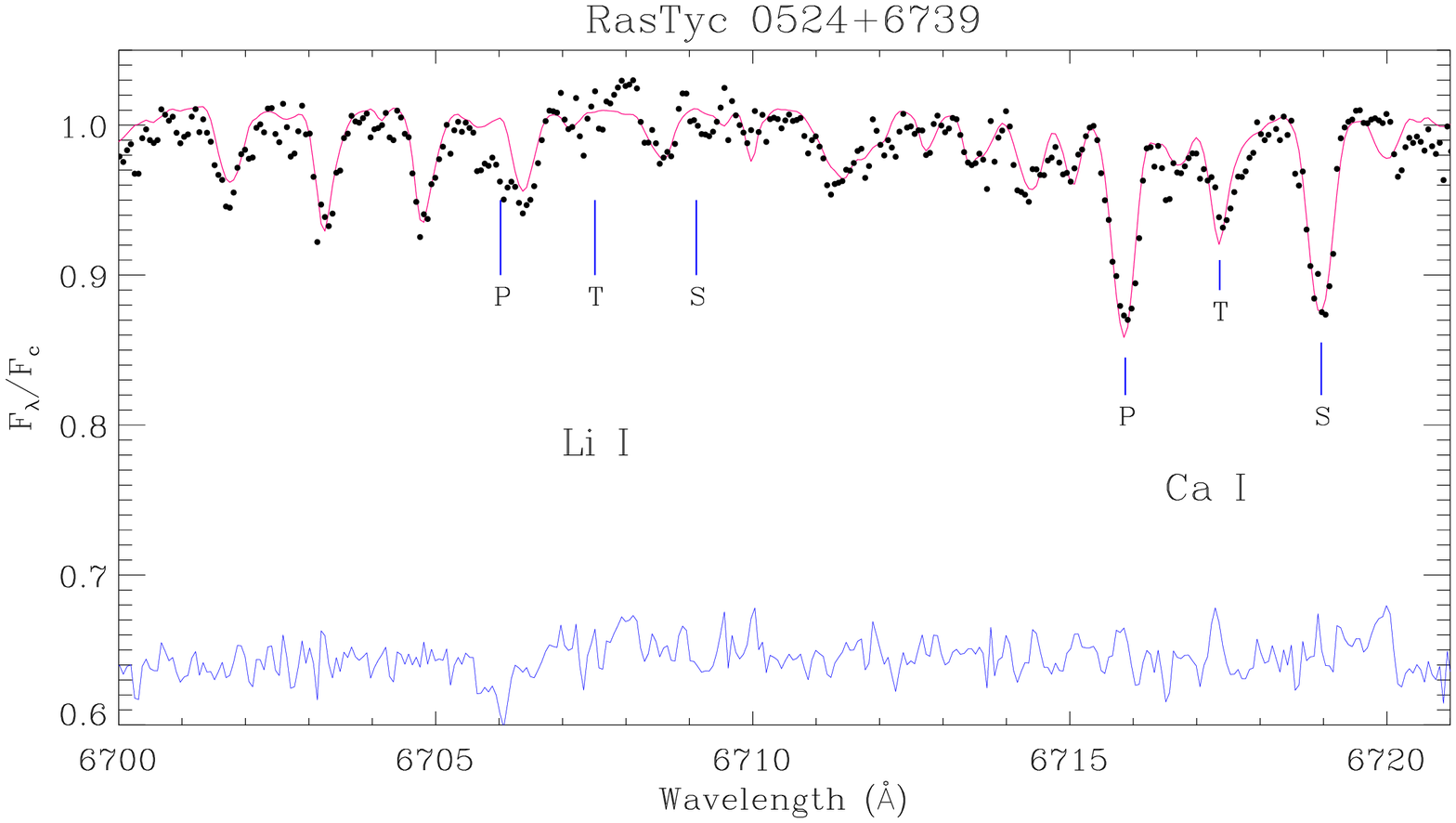}
	\includegraphics[width=8.5cm]{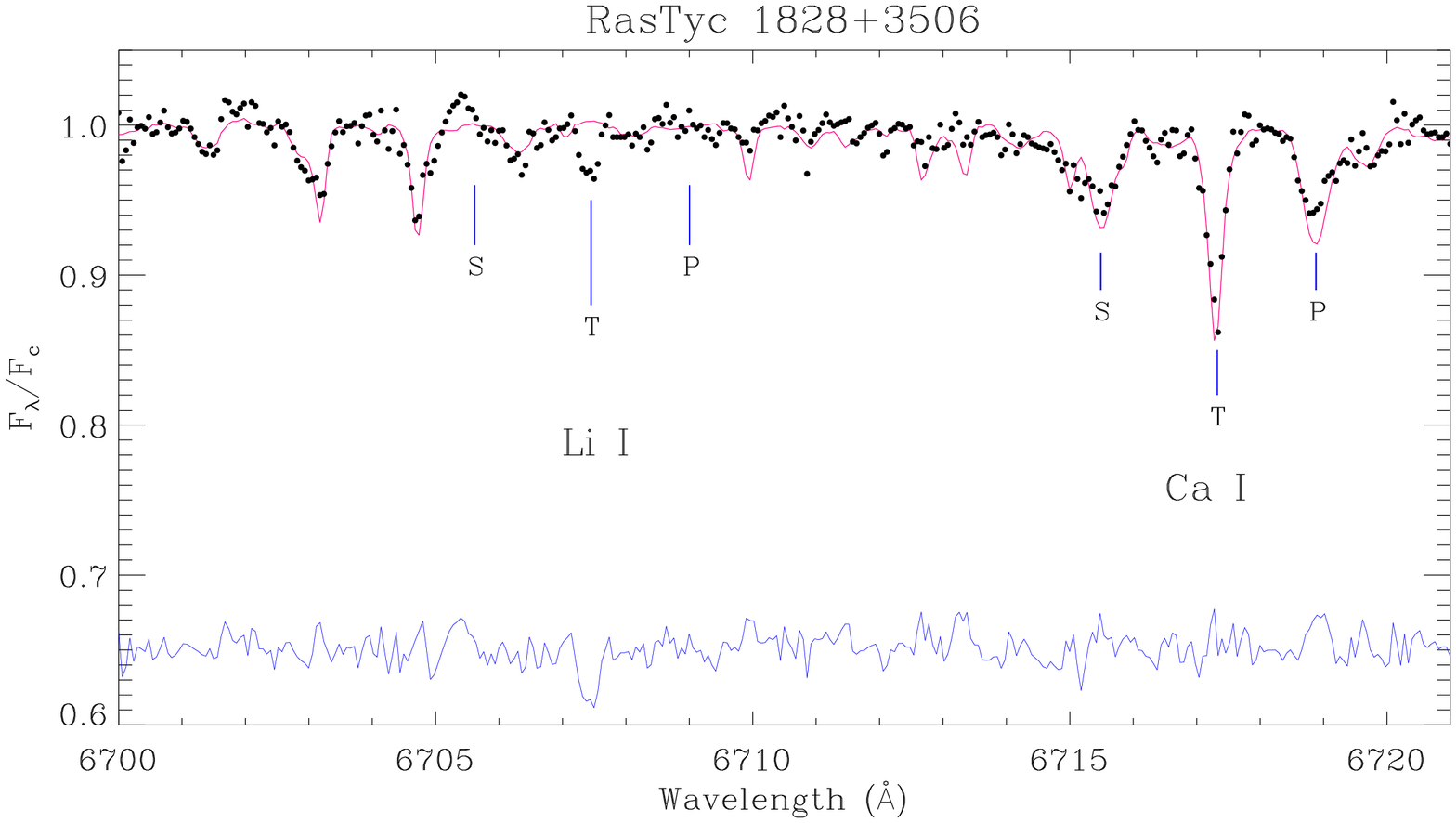}
	\includegraphics[width=8.5cm]{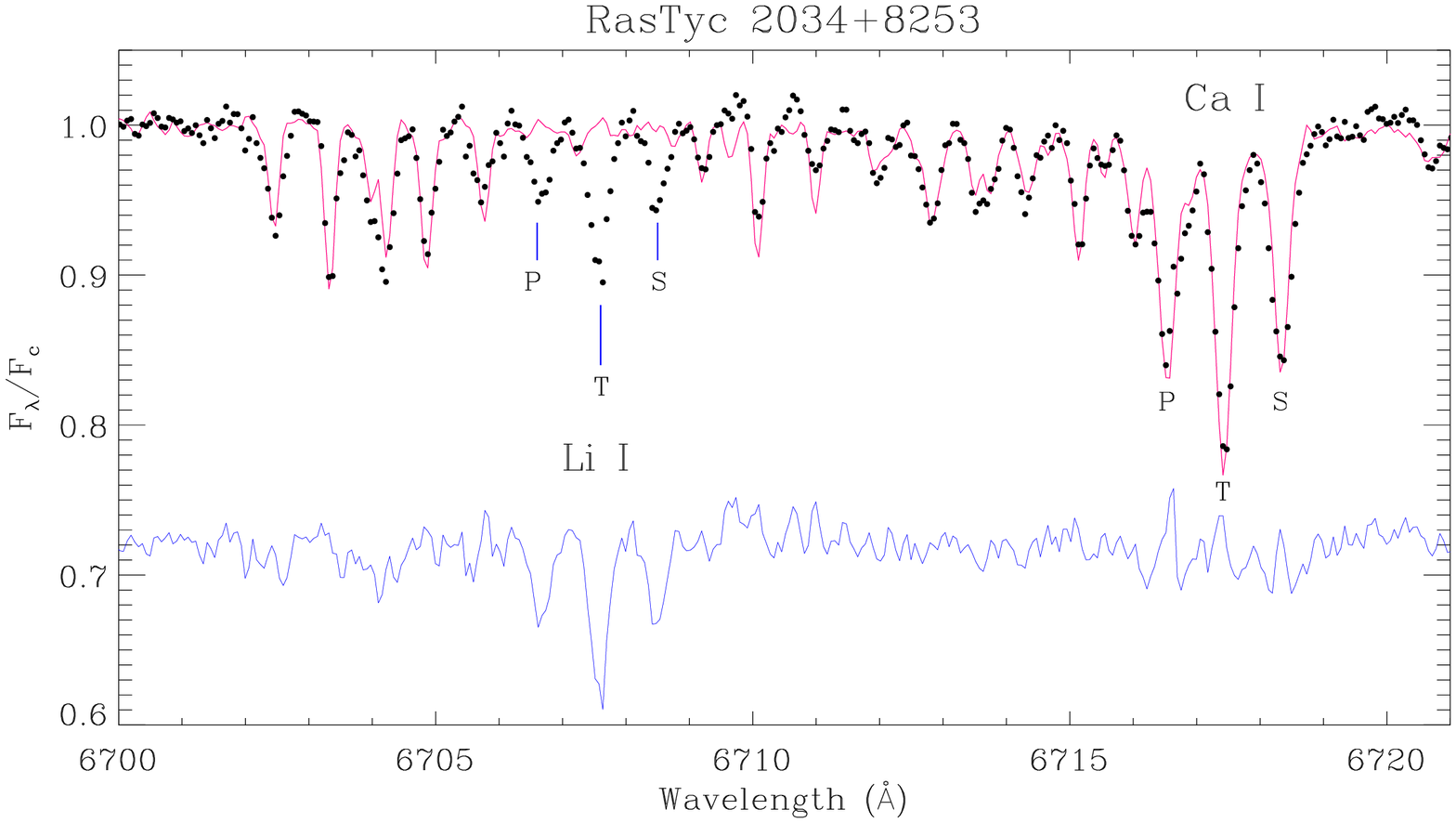}
  \end{center}
	\caption{\label{fig:TripleLi} The spectral region containing the \ion{Li}{i} $\lambda$6707.8 and
	the \ion{Ca}{i} $\lambda$6717.7 lines in a high-resolution spectrum of the three triple systems (dots).
	The reference spectrum built up with the weighted sum of three standard-star spectra 
	(Table~\ref{Tab:ParamSpTComb}) is shown superimposed by a thin line. 
	The Lithium and Calcium lines for the three components of this system are also marked by vertical lines.
	We note the absence of three \ion{Li}{i} absorption lines in each reference spectrum.
	Each box also displays at the bottom the difference (observed - synthetic) with a thin line.}
\end{figure}

The use of ROTFIT and COMPO2 codes allowed us to derive the spectral type and the APs for the components of each system (Table~\ref{Tab:ParamSpTComb}). We used spectra of stars of the same 
spectral types retrieved from the E\textsc{lodie} Archive  \citep{Prugniel01} to build up the reference spectra 
displayed in Fig.~\ref{fig:TripleLi}. 
Moreover, the relative continuum contributions of the tertiary components are in agreement with their 
spectral types found by us, taking into account the errors derived from the distribution of the best spectral 
combinations. This uncertainty is about 1.5 spectral subclasses, except for the tertiary component of 
RasTyc\,0524+6739, the coolest star, for which the uncertainty is more than 2 spectral subclasses. 
Regarding $\log g$ and [Fe/H], despite the rather large errors, we can state that the three systems are composed of MS stars with a nearly solar metallicity.

\begin{table}[t]
\caption{Physical parameters for each component of the three
systems.}
\begin{center}
\begin{tabular}{lccc}
\hline
\hline
\noalign{\medskip}
Component & Primary (P) & Secondary (S) & Tertiary (T) \\
\noalign{\medskip}
\hline
\noalign{\medskip}
\multicolumn{4}{l}{RasTyc\,0524+6739 : }  \\
\noalign{\medskip}
$T_{\rm eff}$ (K) & 5350 $\pm$ 280  &   5270 $\pm$ 270  &   4700 $\pm$ 450 \\
$\log g$     & ~~4.2 $\pm$ 0.3    &    ~~4.2 $\pm$ 0.4        & ~~4.2 $\pm$ 0.4 \\
$[Fe/H]$     & $-0.21 \pm$ 0.18 &     $-0.24 \pm$ 0.21  &    $-0.12 \pm$
0.11 \\
$v\sin i$ (km\,s$^{-1}$)  &  12 $\pm$ 2 & 12 $\pm$ 3 & $<5$ \\
Weight       & ~~0.45 $\pm$ 0.05  &    ~~ 0.43 $\pm$ 0.05 &    ~~0.12 $\pm$ 0.02
\\
Sp. Type     & ~~G9V  & ~~G9V  & ~~K5V \\
\hline
\noalign{\medskip}
\multicolumn{4}{l}{RasTyc\,1828+3506 : }  \\
\noalign{\medskip}
$T_{\rm eff}$ (K) & 5800 $\pm$ 400 &    5800 $\pm$ 350   &  5480 $\pm$ 300 \\
$\log g$     & ~~4.2 $\pm$ 0.2    &    ~~4.2 $\pm$ 0.2        & ~~4.3 $\pm$ 0.2 \\
$[Fe/H]$     & $-0.27 \pm$  0.17 &     $-0.24 \pm$ 0.18   &   $-0.21 \pm$
0.12 \\
$v\sin i$ (km\,s$^{-1}$)  & 12 $\pm$ 1 & 11 $\pm$ 2 & $<5$ \\
Weight       & ~~0.38 $\pm$ 0.08  &    ~~0.33 $\pm$ 0.08 &    ~~0.29 $\pm$ 0.03
\\
Sp. Type     & ~~G1V & ~~G1V & ~~G4V \\
$EW$(Li)  (m\AA)     & --- & --- & ~~49 $\pm$ 15\\
$\log N$(Li)       & --- & --- & ~~1.8 -- 2.0 \\
\hline
\noalign{\medskip}
\multicolumn{4}{l}{RasTyc\,2034+8253 : }  \\
\noalign{\medskip}
$T_{\rm eff}$ (K) & 4960 $\pm$ 260  &   4920 $\pm$ 300   &  5090 $\pm$ 200 \\
$\log g$     & ~~4.3 $\pm$ 0.2     &   ~~4.4 $\pm$ 0.2        & ~~4.4 $\pm$ 0.2 \\
$[Fe/H]$   & $-0.23 \pm$ 0.24   &   $-0.23 \pm$ 0.23   &   $-0.05 \pm$ 0.17
\\
$v\sin i$ (km\,s$^{-1}$)  &  $<5$  & $<5$  & $<5$  \\
Weight       & ~~0.31 $\pm$ 0.04  &     ~~0.30 $\pm$ 0.04 &    ~~0.39 $\pm$ 0.04
\\
Sp. Type     & ~~K3V & ~~K3V & ~~K1V \\
$EW$(Li)  (m\AA)      & ~~72 $\pm$ 20 & ~~76 $\pm$ 10 & ~~96 $\pm$ 14 \\
$\log N$(Li)       & ~~1.9 -- 2.0 & ~~1.9 -- 2.0 & ~~1.9 -- 2.0 \\
\hline
\noalign{\medskip}
\end{tabular}
\end{center}
\label{Tab:ParamSpTComb}
\end{table}

We found that the weight for the components of all inner binaries is 
nearly equal (Table~\ref{Tab:ParamSpTComb}), i.e. their luminosity ratio is $\simeq$\,1. 
This is compatible with the mass ratio (M$_P$/M$_S$)\,$\simeq$\,1 (Table~\ref{Tab:ParamOrb}) derived 
from the solution of the RV curves, if the two {\it twin} stars are both on the MS. Our 
results, although not statistically significant, are in favor of an excess of twins in spectroscopic binaries 
containing a third body as suggested by \citet{Tok06}. 
Moreover, the CCF dips of RasTyc\,1828+3505 and 
RasTyc\,2034+8253 would suggest a tertiary component brighter than each star of the 
inner binary. However, the weights quoted in Table~\ref{Tab:ParamSpTComb} for RasTyc\,1828+3505 are in 
conflict with the depth of the CCF dips. The inconsistency is removed if we take into account the earlier
spectral type and the faster rotation of the components of the inner binary of this system compared to the 
tertiary star. 

We used the nomenclature proposed by \citet{Laf08} assigning the letter A to the brightest (more massive 
for MS stars) component and enclosing in parentheses the components forming the inner binary. We found 
one system (RasTyc\,2034+8253) and two systems (RasTyc\,0524+6739 and RasTyc\,1828+3506) in the 
A,(B,C) and (A,B),C configurations, respectively. Although our sources appear to be older than PMS stage 
(Sect.~\ref{sec:AGE}), the configurations found are similar to those typically encountered in PMS stars 
\citep{Laf08, Correia06}. Moreover, \citet{MaMa87} and \citet{Tok06} found that the most massive component 
in the multiple stellar systems is preferentially in the close binaries. In particular, \citet{Tok06} found a small 
fraction of systems (17 $\pm$ 4\,\%) where the spectroscopic primary is not the most massive star. 

Our results seem to be consistent with those of these authors and need to be confirmed with a larger 
statistical sample of multiple systems. 
The analysis of all the triple systems found by us in the \textsl{RasTyc} sample, for which we are still collecting RV data, will help us to confirm these findings. 

\subsection{Age estimation and kinematics}
\label{sec:AGE}

\begin{table*}
\caption[Kinematic parameters of the new triple systems.]
{Kinematic parameters of the new triple systems.}
\begin{center}
\begin{tabular}{lcccccc}
\hline
\hline
\noalign{\medskip}
Name &  Age & $U$  &  $V$  &  $W$ & Moving Group & Probability \\    
     & (Myr) & (km\,s$^{-1}$) & (km\,s$^{-1}$) & (km\,s$^{-1}$) & & (\%)\\
\noalign{\medskip}
\hline
\noalign{\medskip}
RasTyc\,0524+6739 & $>600$ &    $-16.1  \pm$ 1.9 	&  ~~$-1.1 \pm$ 1.9  &  ~~~0.6 $\pm$ 1.5  & UMa        & ~~$5$\,--\,$35$ \\
RasTyc\,1828+3506 & $400$\,--\,$600 $ &  ~~~8.5 $\pm$ 1.0 &  $-15.6  \pm$ 1.0  &     $-13.7 \pm$ 1.9  & Pleiades & $25$\,--\,$55$ \\
RasTyc\,2034+8253 & $100$\,--\,$300$ &    ~19.1 $\pm$ 4.1 	&  $-14.0 \pm$ 2.0  &   $-17.3 \pm$ 4.8  & IC\,2391  & ~~$5$\,--\,$15$ \\
\hline
\end{tabular}
\end{center}
\label{Tab:Kin}
\end{table*}

It is well established, for stars later than about mid-G spectral type, that the strength of the \ion{Li}{i} 
$\lambda$6707.8~line can be used as an age estimator, a high $\log N$(Li) being a youth indicator. 
Although the Lithium abundance can not be simply converted into age, we can give a rough evaluation 
of the age by comparing the $\log N$(Li) value of our systems (Table~\ref{Tab:ParamSpTComb}) to that of Pleiades and 
Hyades stars having the same temperature \citep[see, e.g.,][]{Sod93, Jef00}. We report the estimated age in 
Table~\ref{Tab:Kin}. 

The parallax ($\pi$) from TYCHO-1 catalog \citep{Hipp} are not enough accurate. Thus, we estimated 
photometric distances (Table~\ref{Tab:PhotoData}) from the ``integrated'' $V$ magnitude measured by 
us and the mean $V$ absolute magnitude for each triple system. The precision of proper motions for 
these systems is $1.5$\,mas\,yr$^{-1}$ in TYCHO-2 catalog \citep{Hog00}. From these two parameters and barycentric 
RVs of the inner binary, we computed the space-velocity 
components ($U,V,W$) of these SB3s in the left-handed coordinate system. 
Their space velocities are consistent with those of the young-disk (YD) population (Fig.~\ref{fig:UV})$^1$. 
Based on two kinematics methods \citep{Klutsch2008}, we determined the membership 
probability (Table~\ref{Tab:Kin}) to five young Stellar Kinematic Groups \citep[SKGs;][]{Montes2001}. 

For RasTyc\,0524+6739, we did not observe any \ion{Li}{i} absorption lines in the spectra (Fig.~\ref{fig:TripleLi}, 
top panel), notwithstanding the spectral types of its components that would permit the detection of the \ion{Li}{i} 
line also with a moderate abundance. Thus, this system should be older than the Hyades. Therefore, 
even though its position in the $UVW$ diagrams points to a marginal association with the young Ursa Major 
(UMa) group ($Age\sim300$\,Myr), we do not consider this star as a new member of this SKG.

Trusting only the kinematics, RasTyc\,1828+3506 could be a new member of the Pleiades moving 
group ($Age\sim100$\,Myr), whose members display strong Lithium absorption. However, this is not
consistent with the non-detection of \ion{Li}{i} absorption lines in the spectra of RasTyc\,1828+3506, except for the tertiary component  
(Fig.~\ref{fig:TripleLi}, middle panel). The $\log N$(Li) value we deduce for it is only slightly higher than 
that of Hyades stars. 
Therefore, the age of the system could be in the range $400-600$\,Myr, ruling out its membership to the 
Pleiades moving group.

Finally, even though the RasTyc\,2034+8253 kinematics is marginally consistent with that of the 
already known members of IC\,2391 supercluster, we can clearly distinguish the Lithium lines for the three 
components (Fig.~\ref{fig:TripleLi}, lower panels). The $\log N$(Li) value found for its three components  
is very similar and reinforce the idea of a common origin for all the components. This value is only slightly 
lower than that of Pleiades stars. Therefore, we estimate an age between 100~and 300\,Myr which is 
compatible with that of two stellar populations in IC\,2391 supercluster \citep{Eggen91}. The agreement 
between the kinematic age and that inferred from the Lithium suggests that this system can be a possible 
new member of this SKG.\\

\onlfig{8}{
\begin{figure*}[!ht] 
	\centering
	\includegraphics[width=9.0cm]{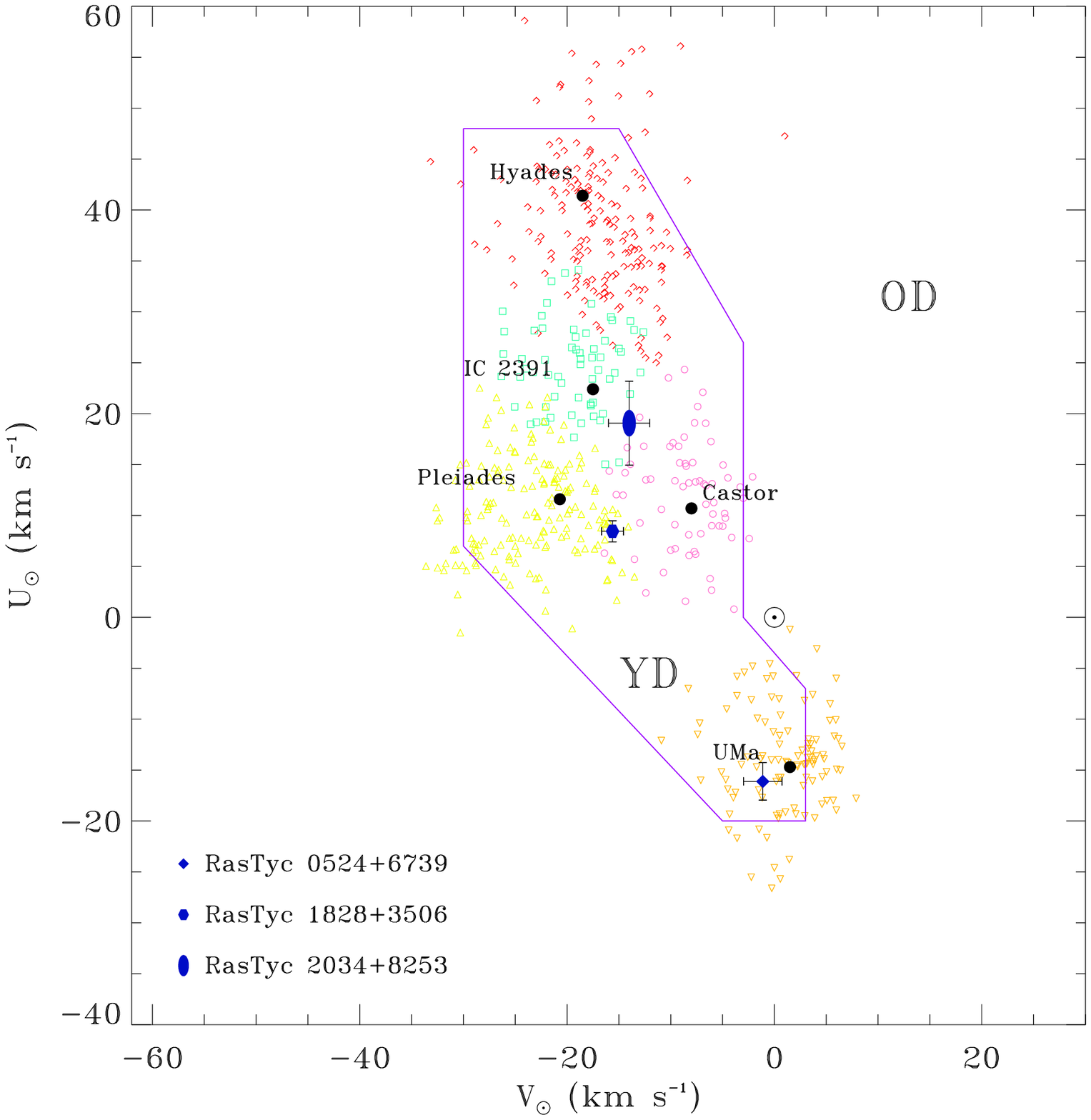}
	\includegraphics[width=9.0cm]{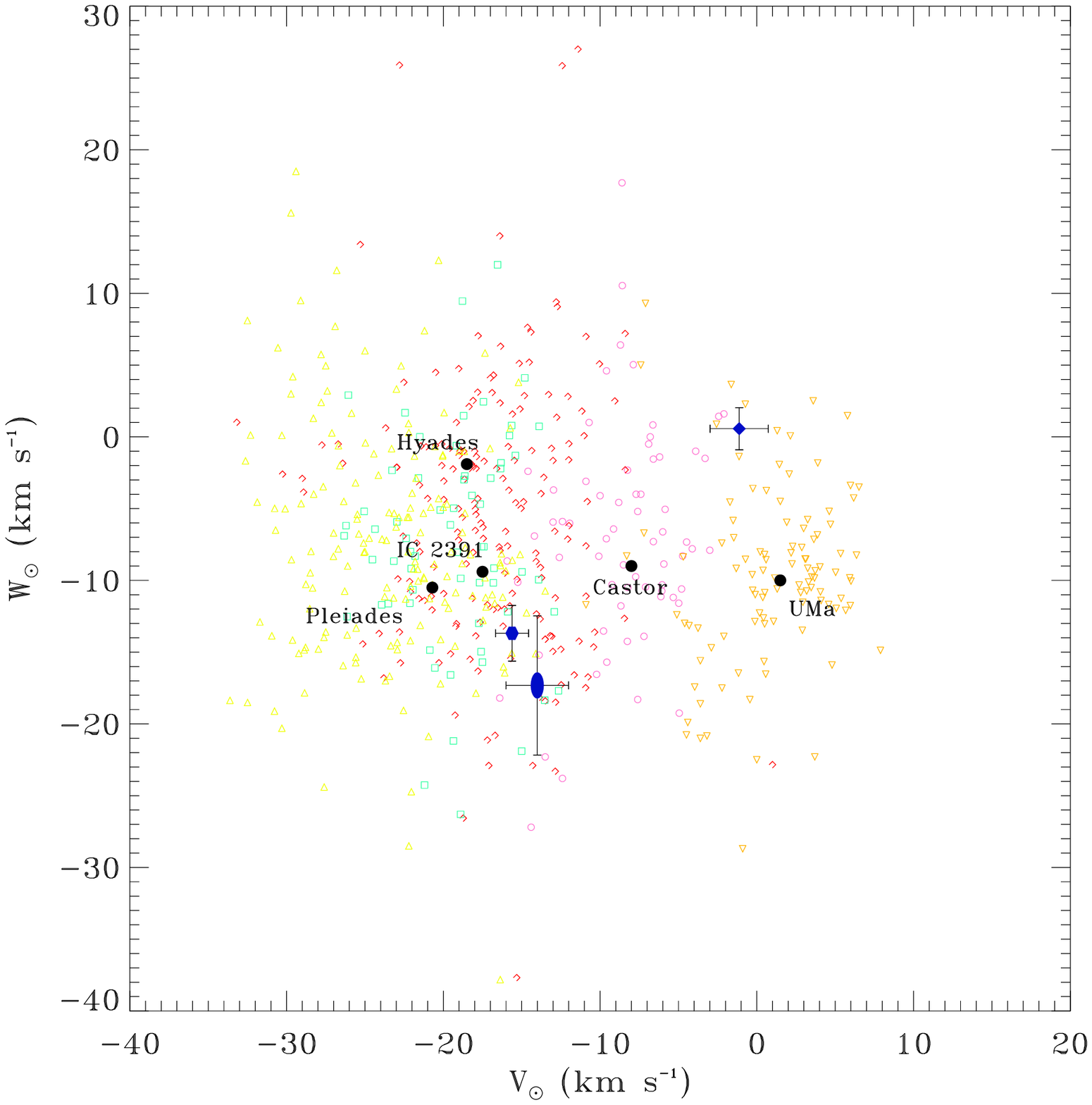}
	\includegraphics[width=9.0cm]{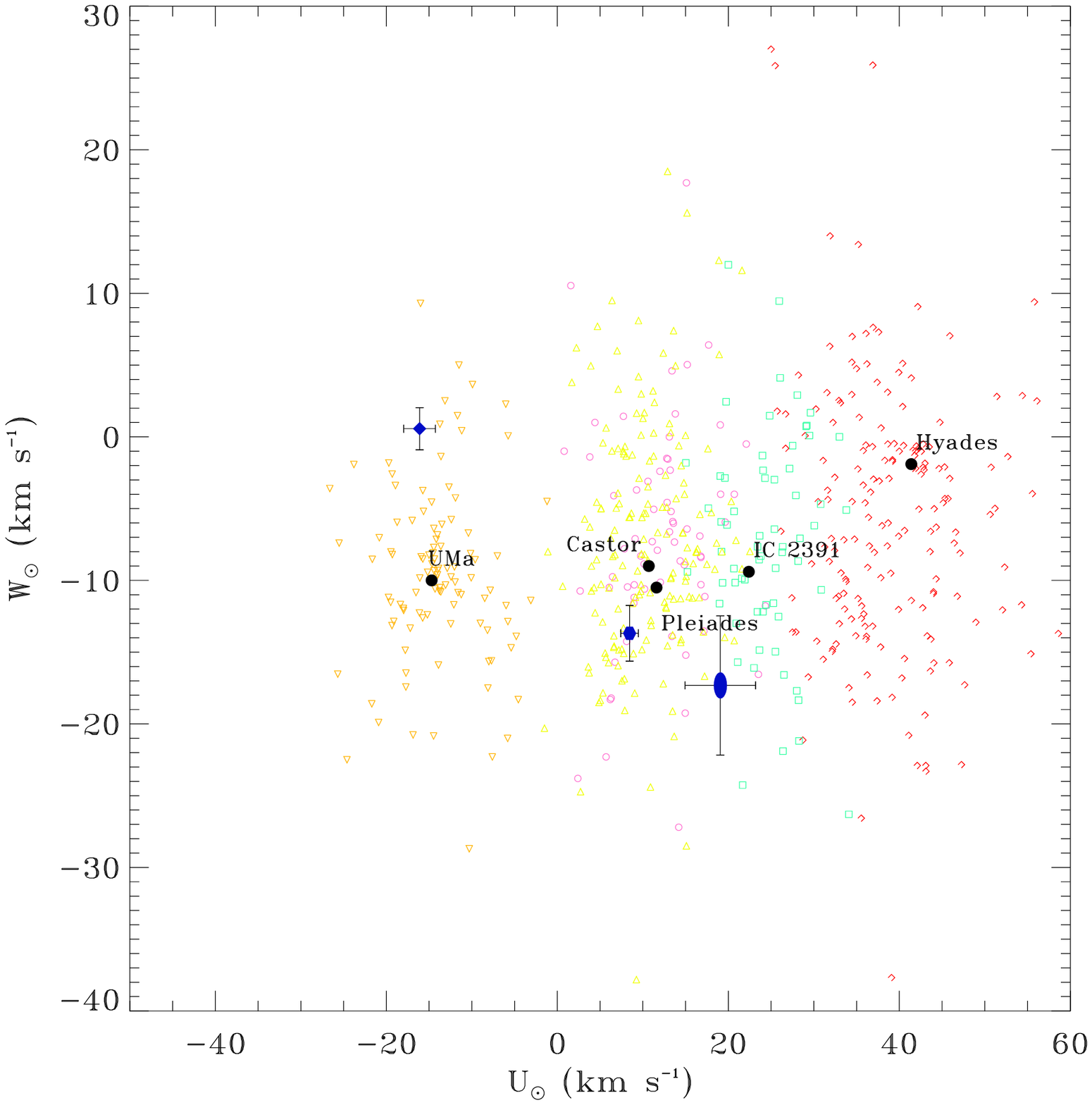}
	\caption{\label{fig:UV} The $U$--$V$ (Left top panel), $V$--$W$ (Right top panel), and $U$--$W$ 
	(Lower panel) diagrams of the {\it RasTyc} triple systems. The average velocity components 
	(dots) of some young SKGs and those of some late-type stars members of these young SKGs 
	\citep{Montes2001} are also plotted (square, triangle, circle, upside down triangle, and U symbols for 
	the IC\,2391 supercluster, Pleiades, Castor, UMa moving groups, and Hyades supercluster, respectively). 
	The locus of the young-disk (YD) and the old-disk (OD) populations \citep{Eggen96} are also marked 
	on the $U$--$V$ diagram.}
\end{figure*}
}

\section{Conclusions}
\label{sec:Conclusions}

This paper is devoted to the analysis of three new triple systems discovered in the \textsl{RasTyc} sample 
of stellar X-ray sources. Their spectroscopic and photometric data allow us to conclude that they are almost 
certainly stable hierarchical triple systems composed of short-period inner binaries plus a tertiary component 
in a long-period orbit. 
The orbital periods of the inner binaries range from 3.5 to 7.6 days and the orbits are practically circular. 
From the high-resolution spectra we 
also found the spectral composition and the astrophysical parameters of the components that turn out to 
be all G-K main sequence stars. In all cases, the components of the inner binaries have nearly the same 
masses, spectral types, and luminosities. From their kinematics and/or Lithium content, these systems result 
to be fairly young. RasTyc\,2034+8253 is the only system in which the \ion{Li}{i}\,$\lambda$6707.8 line is 
strong enough to be clearly visible in the spectra of all the three components and suggests an age in the 
range $100-300$\,Myr. It is a possible new member of the IC\,2391 supercluster. For the remaining systems, 
the membership to young moving groups is rather uncertain. 

Our spectroscopic survey has revealed that multiple systems represent a large fraction of the 
\textsl{RasTyc} sources. However, a detailed analysis is absolutely necessary for drawing statistically 
significant conclusions. Since \textsl{RasTyc} objects are relatively nearby, the discovery and the study 
of new triple systems, such as those presented in the present paper, can contribute to a better 
understanding of the formation and the evolution of close binaries and multiple systems in the solar 
neighborhood. 

\begin{acknowledgements}

We are grateful to the members of the staff of the OHP in conducting our Key Program and those of the 
OAC observatories for their support and help with the observations. 
This research made use of SIMBAD and VIZIER databases, operated at the CDS, Strasbourg, France. 
This publication uses ROSAT data. 
A.~K. also thanks the MEN and ULP for financial support.
A partial support from the Italian {\it Ministero dell'Istruzione, Universit\`a e  Ricerca} (MIUR) is also acknowledged.
\end{acknowledgements}

\Online

\end{document}